\documentclass[aps,pra,twocolumn,showpacs,amsmath,amssymb,superscriptaddress]{revtex4}
\usepackage{graphicx}
\usepackage{subfigure}
\usepackage{color}
\usepackage{array,url}
\usepackage{pdfpages}
\usepackage{rotating}
\usepackage{multirow}
\newcommand{\be}{\begin{equation}}
\newcommand{\ee}{\end{equation}}
\newcommand{\bea}{\begin{eqnarray}}
\newcommand{\eea}{\end{eqnarray}}
\newcommand{\ba}{\begin{array}}
\newcommand{\ea}{\end{array}}

\input{Qcircuit}

\date{\today}

\begin{document}

\title{The two-qubit amplitude damping channel: characterization using quantum stabilizer codes} 
\author{S. Omkar} 
\email{omkar@iisertvm.ac.in}
\affiliation{School of Physics, IISER TVM, CET Campus, Thiruvananthapuram  695016, Kerala, India} 
\author{R. Srikanth} 
\email{srik@poornaprajna.org}
\affiliation{Poornaprajna Institute of Scientific Research, Sadashivnagar, Bengaluru- 560080, India}
\author{Subhashish Banerjee}
\email{subhashish@iitj.ac.in} 
\affiliation{Indian Institute of Technology Rajasthan, Jodhpur- 342011, India} 
\author{Anil Shaji}
\email{shaji@iisertvm.ac.in}
\affiliation{School of Physics, IISER TVM, CET Campus, Thiruvananthapuram  695016, Kerala, India}

\begin{abstract}
A protocol based on quantum error  correction based  characterization of
quantum dynamics (QECCD) is developed for quantum process tomography on a 
two-qubit system interacting dissipatively with a vacuum bath. The method uses a
5-qubit quantum  error correcting code that  corrects arbitrary errors
on the first two qubits, and also saturates the quantum Hamming bound.
The dissipative interaction with a vacuum bath allows for both correlated 
and independent noise on the  two-qubit system. We study the dependence of the degree 
of the correlation of the noise  on evolution time and inter-qubit separation.

\end{abstract}
\maketitle

\section{Introduction \label{sec:intro}}

Quantum  error  correction  (QEC)  is crucial  for  realizing  quantum
information  protocols in  the  laboratory~\cite{Lidar:ts,NC00}. Error
correcting codes are designed  specifically to both detect and correct
for various  types of errors  that may occur during  implementation of
such  protocols. Recently  it  was shown~\cite{OSB15}  that the  error
detection capabilities  which are  an integral part  of the  design of
error correction codes can  be leveraged to perform process tomography
on a register  of qubits. The noise process acting  on the register is
treated  as  the  error and,  in  turn,  error  detection is  done  by
initializing the register in  suitable QEC code states. The statistics
of the errors that are detected during QEC revealing the noise process
is the idea behind  the quantum error correction based characterization of
dynamics (QECCD).

Characterizing the  noise process is  the first step in  combating the
environment  induced loss  of coherence  and entanglement  that almost
always  beset   laboratory  implementations  of   quantum  information
processing  protocols~\cite{Unr1995}. It  can  actually determine  the
strategies  to   be  adopted   towards  the  fault   tolerant  quantum
computation \cite{KLZ1998}.  Such characterization  of the dynamics of
an  open  quantum   system  goes  by  the  name   of  quantum  process
tomography~\cite{NC00}. In  information processing  scenarios, quantum
process  tomography  also  helps   in  benchmarking  the  fidelity  of
implementation of quantum gates \cite{BHP+03} in addition to revealing
the nature of the noise on the system.

A distinct advantage that QECCD  has compared to other quantum process
tomography techniques is  that it can call upon the  extensive body of
literature available  on QEC for  choosing optimal initial  states for
the   register,    useful   measurements   for    syndrome   detection
etc.  Additionally,   once  the  noise  process   is  identified,  QEC
techniques provide the  means to cancel out the  detrimental effect of
the noise as well. The choice of  the type of error correcting code to
be  employed for  process  tomography  depends on  the  nature of  the
physical system  that is being studied  as well as the  type of errors
that are expected.  For instance QEC codes  like CSS codes~\cite{NC00}
and  five  qubit codes~\cite{LMP+96}  are  designed  to perform  error
correction   when   the  noise   on   the   register  of   qubits   is
uncorrelated.  For  correlated  noise models,  under  some  restricted
conditions,  codes  like  the  ones  in  \cite{CAA+11,LNP+11}  may  be
used. In  this Paper  we address the  problem of  characterizing noise
acting on  a pair of  qubits that may or  may not be  correlated using
QECCD.

We completely  characterize the dissipative noise,  due to interaction
with a vacuum bath, on  a two-qubit system \cite{FT02,BRS10} using the
QECCD. This  noise model allows  for both correlated  and uncorrelated
noise  on the  two-qubit system  depending on  the spatial  separation
between them.  Using  QECCD we reconstruct the  quantum process matrix
that describes the noise. The correlated nature of the noise makes the
corresponding process matrix describing the noise non-factorizable.
We also  show  that  the non-classical  correlations characterized
by quantum discord \cite{HV01,OZ02} will be generated between the two qubits due to time evolution 
under a correlated noise even if the qubits are initially in an uncorrelated 
product form.

The rest of this Paper is  structured as follows.  In the next section
we briefly  discuss open  quantum dynamics,  various methods  of doing
process tomography with emphasis on QECCD.  In Sec.~\ref{sec:2AD}, the
dissipative noise on two qubits due  to interaction with a vacuum bath
is described.   We quantify the  degree of correlation  of multi-qubit
noise  as a  departure of  the process  matrix from  product form,  in
Sec.~\ref{sec:D}.  The characterization of the two-qubit noise is done
next in Sec. \ref{sec:characterization}, where all the elements of the
process    matrix,    characterizing    the    noise,    are    worked
out. Sec.~\ref{conclusion} contains a brief discussion of our results.

\section{Open quantum dynamics and QECCD \label{sec:QECCD}}

The noise,  $\mathcal{E}$ acting  on an  open quantum  system starting
initially in  a product state with  its environment is described  by a
completely     positive    (CP),     trace    preserving     dynamical
map~\cite{sudarshan61a}. The  dynamical map on a  quantum state $\rho$
of   dimension   $d$   admits  several   representations.   One   such
representation is in terms of a super-operator:
\begin{equation} 
	\label{eq:Aform}
	\rho' = {\cal E}(\rho) = {\mathcal A} \tilde{\rho}, 
\end{equation}
where $\tilde{\rho}$  is a  $d^{2} \times 1$  dimensional 'vectorized'
version of $\rho$ and ${\mathcal A}$ is a super-operator with a $d^{2}
\times  d^{2}$ matrix  representation.  Alternatively the  map can  be
represented employing an operator sum representation~\cite{kraus83},
\[ \rho'= {\cal E}(\rho) = \sum_{k} K_{j}^{\vphantom{\dagger}} \rho K_{j}^{\dagger}, \qquad  \]
where    the    Kraus    operators,    $K_{j}$    satisfy    $\sum_{j}
K_{j}^{\dagger}K_{j}^{\vphantom{\dagger}}  =  \openone_d$,  for  trace
preserving  map  with  $\openone_d$  being the  identity  operator  of
dimension $d$. The operators $K_j$ can be expanded in terms of a fixed
set of trace-orthonormal operators $F_{l}$, which form a basis for the
set of all operators on the state space of the open system as $K_{j} =
\sum_{l}f_{jl}F_{l}$. The dynamical  map can then be  expressed in the
fixed basis as
\begin{equation}
	\label{eq:processmatrix}
	 \rho'= {\cal E}(\rho) = \sum \chi_{l,m} F_{l}^{\vphantom{\dagger}} \rho F_{m}^{\dagger}, 
\end{equation}
with
\[ \qquad \chi_{l,m} = \sum_{j} f_{jl}f_{jm}^{*} . \]
The dimensions  of the  process matrix  $\chi_{l,m}$ is  $d^{2} \times
d^{2}$. However, since  ${\mathcal E}$ is trace  preserving, there are
$d^{2}$  constraints  on  $\chi_{l,m}$ given  by  $\sum_{lm}\chi_{l,m}
F^\dagger_l   F_m^{\vphantom{\dagger}}  =   \openone_d$,  leading   to
$d^4-d^2$ independent  real elements for  the process matrix.  In this
work the  basis $\{F_l\}$ has  multi-qubit Pauli operators  (string of
$X,~Y,~Z$ and $\openone_2$)  as the elements and is known  as the {\it
  error basis} being appropriate for employing the QEC formalism.

In quantum process tomography, the dynamical map is reconstructed from
observations by  either obtaining the  elements of $\chi_{l,m}$  or by
obtaining   the  Kraus   operators  or   else,  in   some  cases,   by
reconstructing the  matrix representing the super-operator  $A$. Other
useful  representations of  the map  like the  operator sum-difference
representation (OSDR) also exist  \cite{OSB15/1} . In standard quantum
process  tomography~\cite{NC00,DAr00},  a  set  of  suitably  prepared
states   $\{\rho_i\}$  are   input  to   an  unknown   noisy  dynamics
$\mathcal{E}$ to  be characterized and the  corresponding final states
$\rho_{f}$ are measured using state tomography. Repeating this process
for a sufficiently  large number of linearly  independent input states
$\rho_{i}$ the elements of $\chi_{l,m}$ can be obtained. Typically the
number of measurements  required to reconstruct the  process scales as
$d^{4}$  where  $d$  is  the   dimension  of  the  quantum  system  of
interest. Note that typically the system  of interest is a register of
$n$ qubits  whose Hilbert  space dimension scales  as $2^{n}$  and for
such  systems,   the  resources  required  to   perform  full  process
tomography scales exponentially with $n$ as $2^{4n}$.

There are  several refinements on standard  quantum process tomography
with   more  favourable   scalings   of  the   resources  (number   of
measurements)  required to  reconstruct  a dynamical  map. In  ancilla
assisted quantum process  tomography, an ancillary system  is added to
the quantum system  of interest and initialized  in possibly entangled
states with  again possibly non-separable measurements  on both system
and  ancilla following  the  action  of the  unknown  dynamics on  the
system. Reconstruction  of $\chi_{l,m}$ is then  possible with $d^{2}$
measurements~\cite{DP01,ABJ+03,Dar02}.   Direct  characterization   of
quantum dynamics~\cite{ML06,ML07} bypasses the many instances of state
tomography needed in standard  and ancilla assisted process tomography
and  the  complete process  matrix  can  be determined  using  $d^{2}$
different  input   states  and   only  single  measurements   on  each
state. Other  techniques includes characterization of  the noise using
an efficient method for transforming a quantum channel or process into
a  symmetric channel  having  only diagonal  elements  in the  process
matrix   via  twirling~\cite{ESM+07,SMK+08}.   A  method   similar  to
\cite{ESM+07}, but  extended to estimate any  given off-diagonal term,
was introduced in~\cite{BPP08} and it  was further extended to perform
process tomography with out using any ancilla in~\cite{SBL+11}.

\subsection{QECCD }

In  QECCD, the  system of  interest  is a  register of  $p$ qubits.  A
register of $n$ qubits ($n>p$) is initialized into a QEC code state of
the form
\begin{equation}
|\Psi_L\rangle \equiv \sum_{j=0}^{2^k-1} \beta_j|j_L\rangle,
\end{equation}
where $\{|j_L\rangle\}$ denotes a logical  basis for the code space of
a $[[n,k]]$ QEC code which encodes  $k$ qubits into $n$. The first $p$
qubits of the quantum register form the principal system \textbf{P} of
dimension $d=2^{p}$ and the remaining $q = n-p$ qubits form an ancilla
\textbf{A}.

The size  of the ancilla, $q$,  is determined by the  error correcting
properties  of the  particular QEC  code that  is employed  and it  is
chosen such  that all the  $4^p = d^{2}$  errors on \textbf{P}  can be
detected and corrected. Hence the  quantum register should satisfy the
Hamming bound \cite{Got09}
\begin{equation}
	2^k4^p\leq2^{p+q},
	\label{eq:ham}
\end{equation} 
from which we have $q\geq p+k$. For the case where the bound in (\ref{eq:ham}) is saturated and $k=1$ we have $q=p+1$. 
%The size of the error ball, here the set of all correctable errors on $\mathbf{P}$, is $4^{p}$. 
We have made the assumption that there are no appreciable errors on ancillary qubits. However, the assumption can be relaxed using the ambiguous quantum error correcting codes~\cite{OSB15/2} at the cost  of having to do multiple initial state preparations of the $n$ qubit register. 

The  underlying open  quantum dynamics  of $\mathbf{P}$  determine the
statistics of errors on the $n$  qubit register. Note that we are, for
simplicity,  assuming that  the noise  alone  is the  dynamics of  the
register  and deterministic  evolution,  if any,  of  the register  is
either  trivial  or  alternatively,  the   discussion  is  set  in  an
appropriate frame in which such evolution is trivial. While the choice
of the QEC  code determines the `errors' that can  be detected, we use
the stabilizer formalism \cite{Got1997, Got09, NC00} for measuring the
error statistics thereby revealing the  elements of the process matrix
$\chi$. The  stabilizers $S_j$ are  a set of $n-k$  mutually commuting
binary $n$-qubit Pauli observables that stabilize the code space i.e.,
\begin{equation}
	S_j|j_L\rangle = |j_L\rangle.
\end{equation}
 
Any error given by the action of the operator $F_{j}$ on the qubit register can be detected and hence  corrected if for any pair $F_j, F_k$ ($j\ne k$), there exists at least one $S_l$ that anti-commutes with the product $F_jF_k$. This ensures that the set of measurement outcomes of the stabilizers $S_l$'s, collectively forming the error syndrome, will be distinct for every pair $F_j$ and $F_k$ leading to unambiguous error detection. The stabilizer measurement collapses the noisy state into the pure state $|\Psi^{x}_{L} \rangle \equiv F_x|\Psi_L\rangle$, where $F_{x}$ is the error that is detected and this state can be corrected by applying $F^\dag_x=F_x^{\vphantom{\dagger}}$ to bring back the QEC code to its initial state. 

\subsubsection{Diagonal terms of $\chi$.\label{sec:diag}}
The probability to obtain syndrome $x$ corresponding to a error $F_x$ on the quantum register is
\begin{eqnarray} 
	\xi(x) & = &  \textrm{Tr}\left(\mathcal{E}\left( |\Psi_L\rangle\langle\Psi_L|\right) |\Psi^x_{L}\rangle\langle\Psi^x_{L}|\right) \nonumber \\
	 &= &   \langle\Psi_L^x| \left[ \sum_{l,m=0}^{d^2-1}\chi_{l,m}  |\Psi_L^l\rangle\langle \Psi_L^m| \right] |\Psi_L^x\rangle \nonumber \\
	& = &  \sum_{l,m=0}^{d^2-1}  \chi_{l,m} \delta_{x,l} \delta_{x,m} = \chi_{x,x}.
	\label{eq:diag}
\end{eqnarray} 

From Eq.~(\ref{eq:diag}) it is clear  that the diagonal element of the
process matrix $\chi_{x,x}$ is nothing  but the probability with which
error $F_x$ occurs. The error  statistics and the probability for each
detectable  error  characterized  by  syndrome $x$  can  be  found  by
repeating   the  stabilizer   measurements  on   the  register   after
$\mathbf{P}$ is subject to the  noise. These measurements, done on the
register a  fixed number of  times depending on the  desired accuracy,
yields  all  the   $d^{2}$  diagonal  terms  of   the  process  matrix
corresponding to unknown dynamics $\mathcal{E}$. Since all the $S_j$'s
are mutually  commuting it is  possible to obtain all  the measurement
outcomes simultaneously on a  single state preparation. Every syndrome
measurement must be followed by  appropriate error correction steps if
one is to bring the quantum register to its initial state.
 
\subsubsection{Off-diagonal terms of $\chi$.\label{sec:offdiag}}

To determine the $d^4-2d^2+1$ independent off-diagonal terms of a process matrix one has to pre-process the noisy quantum register with a suitable unitary operator $U(a,b)$ after the action of the noise, but before measuring the stabilizers. It is to be noted that pre-processing with any unitary rotates the set of correctable states to another set of correctable states without altering the error correction capability of the QEC code~\cite{OSB15}. This means that the syndrome detection using stabilizer measurements and error correction can be performed as described earlier in Sec.~\ref{sec:diag}.

Now consider the unitary operators,
\bea 
	U(a,b) &= \frac{F_a + F_b}{\sqrt{2}}  \quad \textrm{if} \quad & \{F_a, \, F_b\}= 0 \nonumber\\
	U(a,b) &= \frac{F_a + iF_b}{\sqrt{2}} \quad \textrm{if} \quad & [F_a, \, F_b]=0\label{eq:uab}
\eea
such that $F_aF_x$ and $F_bF_x$ represent correctable errors on \textbf{P} of the quantum  register. Let $g_AF_A\equiv F_aF_x$, where $F_A$ is a Pauli group operator and the \textit{Pauli factor}  $g_A \in\{\pm1,\pm i\}$. Similarly, let $g_BF_B \equiv F_bF_x$.

The probability that the stabilizer measurements find error syndrome $x$ on the quantum register preprocessed, prior to measurement, with the unitary operator $U(a,b)=(F_a +F_b)/\sqrt{2}$ is
\begin{eqnarray} 
 	\xi(a,b,x) &\equiv& \textrm{Tr}\left(U(a,b)\mathcal{E}\left( |\Psi_L\rangle\langle\Psi_L|\right)(U(a,b))^\dag
 |\Psi^x_{L}\rangle\langle  \Psi^x_{L}|\right) \nonumber \\
 	&=& \frac{\chi_{A,A} + \chi_{B,B}}{2} + \frac{g_A^\ast g_B \chi_{A,B} + g_A g_B^\ast\chi_{B,A}}{2}.
	\label{eq:offdiag}
\end{eqnarray}
Both $\chi_{A,A}$ and $\chi_{B,B}$ are known from the error statistics obtained without pre-processing. Eq. (\ref{eq:offdiag}) shows that preprocessing the quantum register with $U(a,b)$ alters the syndrome statistics in a manner that reveals the off-diagonal terms of the process matrix. Repeated measurement of stabilizers followed by error correction repeated sufficient number of times would determine the off-diagonal terms of the process matrix $\chi$ as detailed in~\cite{OSB15}.   When $[F_a,F_b]=0$, we have $U(a,b)= (F_a +iF_b)/\sqrt{2}$ and  in place of Eq. (\ref{eq:offdiag}) we obtain:
\begin{equation}
	\xi(a,b,x) = \frac{1}{2}(\chi_{A,A} + \chi_{B,B} + i\left[g_A g_B^\ast \chi_{B,A} - g_A^\ast g_B\chi_{A,B}\right]).
\label{eq:offdiag+}
\end{equation}

\subsubsection{Toggling\label{sec:tog}}
Depending on the Pauli factors $g_{A}$ and $g_{b}$ appearing in Eqs.~(\ref{eq:offdiag}) and (\ref{eq:offdiag+}) either the real or imaginary parts of each of the off-diagonal terms of $\chi$ would remain undetermined~\cite{OSB15}. We can solve this problem by further pre-processing the noisy state prior to the application of $U(a,b)$ with the operator $T^+ = T \oplus I^\prime$. In $T^{+}$, the operator
\begin{equation}
	T \equiv \sum_{m=0}^{d^{2}-1} e^{i\theta_m}\Pi_L^m,
\label{eq:toggle}
\end{equation}
 acts on all the states within the error ball of the QEC code with  $\Pi_L^m$ being the projectors on to the erroneous logical space given by $F_m\Pi_{\mathcal{C}}F_m$, where $\Pi_{\mathcal{C}}$ is the set of all codewords of the QEC scheme. The unit operator $I'$ acts on the space of states lying outside the error ball. The angles $\theta_m$ are chosen from the set $\{\pm  \pi/4\}$, with equal number of entries of each sign appearing in $T$. As shown in~\cite{OSB15} this `toggling' operation interchanges the real and imaginary parts of $\chi_{l,m}$ prior to $U$, depending on the choice of values of $\theta_{l}$ and $\theta_{m}$. The error correction capability of the QEC code is not affected by the application of the toggling operation $T^+$.  In practical terms, $T^+$ represents $2^k$ copies of the usual $T$-gates,
 \[ T = \frac{1}{\sqrt{2}} \left( \begin{array}{cc}
 	1 + i & 0 \\ 0 & 1-i \end{array} \right). \]
 in the $l{-}m$ subspace of the error ball of each codeword. After identifying the real or imaginary parts of $\chi_{l,m}$ that were not determined by the application of $U(a,b)$, suitable toggling operations can be inserted prior to the application of the unitary so as to exchange the real and imaginary parts of the $\chi_{l,m}$'s of interest and the same procedure followed earlier can be used to estimate these as well. 

With one experimental configuration i.e., preprocessing the quantum register with a $U(a,b)$ and followed by stabilizer measurements one can determine $d^2/2$ off-diagonal elements out of the $(d^4-2d^2+1)/2$ real  parameters to be estimated. Thus $(d^4-2d^2+1)/d^2$ configurations are needed to obtain all the off-diagonal elements of $\chi$ and the total number of configurations needed to completely characterize $\chi$ becomes $2(d^2-1)$.

\section{Two-qubit amplitude damping (2AD) error model \label{sec:2AD}}

The open quantum system we focus on consists of two qubits formed by a pair of atoms with two addressable and distinct levels of interest. The energy gap between the levels $|g_{j} \rangle$ ($j=1,2$) and the corresponding `excited state' $|e_{j}\rangle$ is $\hbar \omega_{j}$. Each atom has a dipole moment $\vec{\mu}_{j}$ and they are respectively located at $\vec{r}_{j}$ with the inter-atomic separation given by $r_{12} = |\vec{r}_{1} - \vec{r}_{2}|$. The two atoms are sitting in a bath of electromagnetic radiation which forms the environment of the open system. The interaction between the qubits and the bath is position dependent and it depends on the dipole moments of the atoms as well. The Hamiltonian for the system, employing the dipole approximation for the qubit bath interaction is~\cite{FT02}
\begin{eqnarray}
	H&=&H_S+H_B+H_{SR}\nonumber\\ 
	&=&\frac{1}{2}\sum_{j=1,2}\hbar\omega_j Z_{j} + \sum_{\vec{k}, s}\hbar\omega_k(b_{\vec{k}s}^\dagger b_{\vec{k}s}^{\vphantom{\dagger}} +1/2)\nonumber\\
	&&-i\hbar\sum_{ks}\sum_{j=1,2}\vec{\mu}_i.\vec{g}_{\vec{k}s}(\vec{r}_j)(S_j^+
+ S_j^-)b_{\vec{k}s}-h.c, 
\end{eqnarray}
where  $S_j^{\pm}= (X_{j} + iY_{j})/2$ are the rising and lowering operators on $j^{\rm th}$ qubit, $b_{\vec{k}s}$ is the annihilation operator corresponding to a bath mode $\vec{k}s$, with wave vector $\vec{k}$, frequency $\omega_k$ and polarization index $s=1,2$. Here we are labelling the three Pauli operators on each of the qubits as $X_{j}$, $Y_{j}$ and $Z_{j}$ respectively. The system-bath coupling has the form 
\begin{equation}
	\vec{g}_{\vec{k}s}(\vec{r}_j)=\sqrt{\frac{\omega_k}{2\epsilon\hbar V}} \vec{\varepsilon}_{\vec{k}s} e^{i \vec{k} \cdot \vec{r}_j},
\end{equation}
which is also the mode function of the bath evaluated at position $\vec{r}_j$ of the $j^{\rm th}$ qubit and $\vec{\varepsilon}_{\vec{k}s}$ is the unit polarization vector of the bath mode with $V$ as the normalization volume.

This model was analyzed in detail in~\cite{FT02} where using the Born-Markov and rotating wave approximations, the following master equation is obtained for the density matrix of the two qubit system when it is interacting with a zero temperature electromagnetic bath with no squeezing:
\begin{eqnarray}
	\label{eq:master1}
	\frac{d\rho}{dt} & = & -\frac{i}{\hbar}[H_{D}, \, \rho]  \nonumber \\
	&& \; - \; \frac{1}{2} \sum_{j,k=1}^{2} \Gamma_{jk}  (\rho S_{j}^{+} S_{k}^{-} + S_{j}^{+} S_{k}^{-} - 2 S_{j}^{-} \rho S_{j}^{+}), \quad
\end{eqnarray}
where 
\begin{equation}
	H_{D}=\frac{1}{2}\sum_{i=1,2}\hbar\omega_j Z_{j} + \hbar\sum_{j\neq k}\Omega_{jk} S_j^+S_k^-.
\end{equation}
It is convenient to express the reduced density matrix of the system in the `dressed state' basis which is the eigenbasis of the Hamiltonian $H_{D}$ given above. When the two qubits are identical, the dressed state basis is given by:
\begin{eqnarray}
	|g \rangle = |g_1 \rangle |g_2 \rangle, \qquad & |s \rangle = \frac{1}{\sqrt{2}} (|e_1 \rangle |g_2 \rangle +  |g_1 \rangle |e_2 \rangle), \nonumber\\
	|e \rangle = |e_1 \rangle |e_2 \rangle, \qquad &   |a \rangle = \frac{1}{\sqrt{2}} (|e_1 \rangle |g_2 \rangle -  |g_1 \rangle |e_2 \rangle).
	\label{dressed}
\end{eqnarray}
with the corresponding eigenvalues being $E_g = -\hbar \omega_0$, $E_s = \hbar \Omega_{12}$, $E_a = -\hbar \Omega_{12}$ and $E_e = \hbar \omega_0$. 

The Hamiltonian $H_{D}$ describes the dynamics in the presence of the vacuum induced coherent dipole-dipole interaction between the two qubits. The strength of this interaction is given by $\Omega_{jk}$. For the two qubit case, we have
\begin{equation}
	\label{eq:omega12}
	\Omega_{12} = \Omega_{21} = \frac{3}{4}\sqrt{\Gamma_{11} \Gamma_{22}} {\mathcal G}( r_{12}),
\end{equation}
with
\begin{eqnarray*}
	 {\mathcal G}(r_{12}) & = &  -[1- (\hat{\mu} \cdot \hat{r}_{12})^2]\frac{\cos(k_0 r_{12})}{k_0 r_{12}}  \nonumber \\
	 && \;  + \,  [1-3(\hat{\mu} \cdot\hat{r}_{12})^2] \bigg[\frac{\sin(k_0 r_{12})}{(k_0 r_{12})^2} + \frac{\cos(k_0 r_{12})}{(k_0 r_{12})^3} \bigg], \quad
\end{eqnarray*}
where $\hat{\mu} = \hat{\mu}_{1} = \hat{\mu}_{2}$ is the unit vector along the electric dipole moment of the atomic qubits which are assumed to be aligned parallel to each other and $\hat{r}_{12}$ is the unit vector along the line joining the two qubits. The $\Omega_{12}$ term in $H_{D}$ is responsible for coherently driving the excited state population between the two atomic qubits. The $\Gamma_{jj}$ appearing in Eq.~(\ref{eq:omega12}), are the spontaneous emission rates of each of the qubits due to their independent dissipative interactions with the bath and these rates are given by 
\[ \Gamma_{jj} =\frac{\omega^3_j \mu^2_j}{3\pi\epsilon \hbar c^3}.\]
For the identical qubit case, $\Gamma_{11} =\Gamma_{22} = \Gamma$. 

The coefficients $\Gamma_{jk}$ for $j \neq k$, appearing in the master equation (\ref{eq:master1}) represent collective, incoherent dissipation rates due to the system bath interaction. For the two qubit system we have
\begin{equation}
	\label{gamma} 
	 \Gamma_{12} =  \Gamma_{21} = \frac{3}{2}\sqrt{\Gamma_{11} \Gamma_{22}} {\mathcal F}(r_{12}) = \frac{3}{2} \Gamma {\mathcal F}(r_{12}), 
\end{equation}
where
\begin{eqnarray*}
	{\mathcal F}(r_{12}) & =&  [1-(\hat{\mu}.\hat{r}_{12})^2]\frac{\sin(k_0 r_{12})}{k_0 r_{12}} \nonumber\\ 
	&& \; + \;  [1-3(\hat{\mu}.\hat{r}_{12})^2] \bigg[ \frac{\cos(k_0 r_{12})}{(k_0 r_{12})^2}-\frac{\sin(k_0 r_{12})}{(k_0 r_{12})^3} \bigg].\nonumber\\
\end{eqnarray*}

\begin{figure}
	\resizebox{8cm}{5cm}{\includegraphics{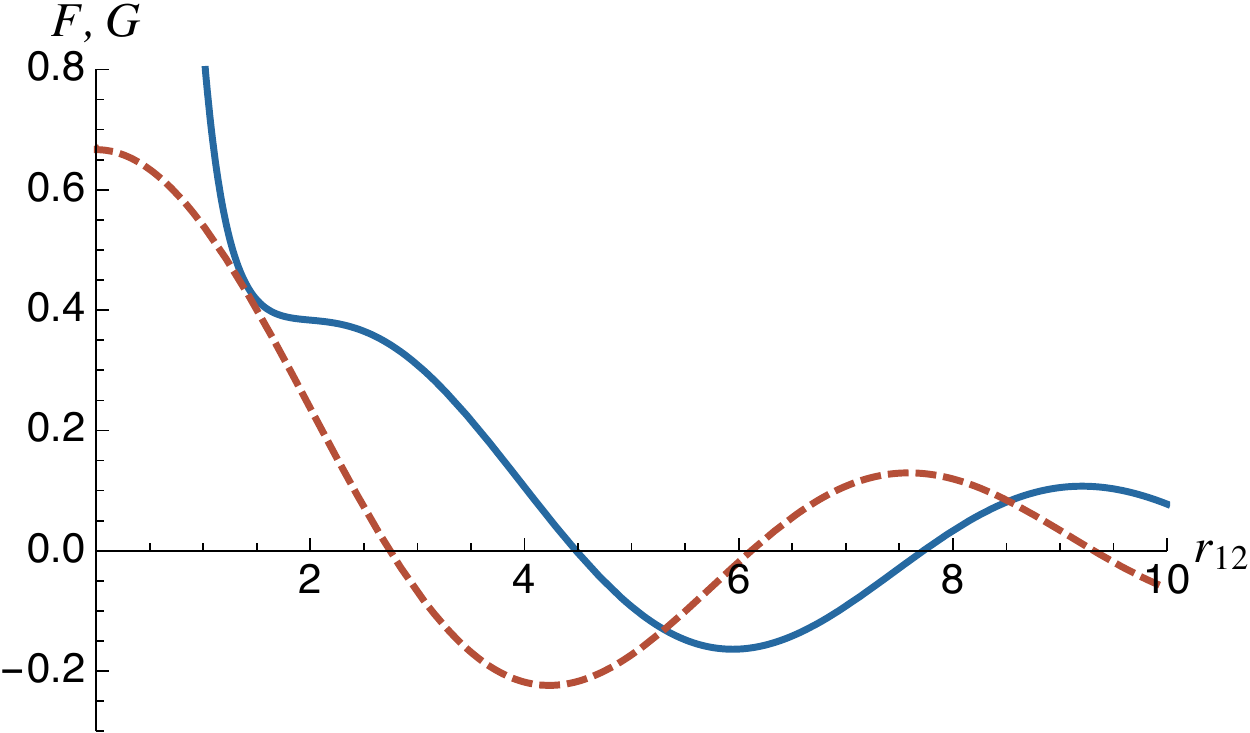}}
	\caption{The  functions  ${\mathcal F}(r_{12})$  (dashed,  red
          line) and ${\mathcal G}(r_{12})$  (blue line) are plotted as
          a function of the separation between the qubits in arbitrary
          units when  the dipole  moments of  both qubits  are aligned
          perpendicular  to   the  line  joining  the   two  and  with
          $k_{0}=1$. \label{fig1}}
\end{figure}

The  functions  ${\mathcal   G}(r_{12})$  and  ${\mathcal  F}(r_{12})$
represent  the  spatial  dependence  of the  coherent  and  incoherent
interactions between the qubits  mediated by the bath. The wavelength,
$\lambda_{0} =  2\pi/k_{0}$ corresponding  to the wave  number $k_{0}$
appearing in the expressions for ${\mathcal G}(r_{12})$ and ${\mathcal
  F}(r_{12})$ sets the length  scale for such interactions between the
two qubits.  Note that $k_{0}$  is also equal to  $\omega_{0}/c$ where
$\omega_{0} = (\omega_{1} +  \omega_{2})/2$ and hence $\lambda_{0}$ is
the resonance  wavelength of the  qubits when they are  identical. The
two functions are plotted in Fig.~\ref{fig1}

From Fig.~\ref{fig1} we see that  as the separation between the qubits
$k_{0}r_{12}   \rightarrow  0$  the   functions  ${\mathcal   F}$  and
${\mathcal G}$ increase. The  bath mediated coherent dynamics governed
by $\Omega_{12}$ as  well as the collective decoherence  rate given by
$\Gamma_{12}$ both  become comparable  to the independent  dynamics of
the  qubits  and  we   obtain  a  collective  decoherence  model  with
predominantly correlated errors. On the other hand $k_{0}r_{12} \gg 1$
we have  an independent decoherence model  with the errors  due to the
influence of the  bath on each qubit being  acting independent of each
other.

We assume that the initial state of the system and environment is a product state so as to ensure completely positive reduced dynamics. Let the initial state of the environment be the vacuum and that of the two-qubit system in the dressed state basis be 
\begin{equation}
	\label{eq:inistate}
	\rho= \left( \begin{array}{clclr}
		\rho_{ee}&\rho_{es}&\rho_{ea}&\rho_{eg}\\
		\rho_{se}&\rho_{ss}&\rho_{sa}&\rho_{sg}\\
		\rho_{ae} & \rho_{as}&\rho_{aa}&\rho_{ag}\\
		\rho_{ge} & \rho_{gs}&\rho_{ga}&\rho_{gg}
	\end{array} \right).
\end{equation} 
After the pair of qubits and its environment has interacted for a duration $t$, the time evolved reduced state of the system is given        by \cite{BRS10}
\begin{eqnarray}
&&\rho_{\mathcal{E}}= {\mathcal E}^{\rm 2AD} (\rho) = \nonumber\\
&&\left( \begin{array}{cccc}
A\rho_{ee}&J\rho_{es}&M\rho_{ea}&L\rho_{eg}\\
J^\ast\rho_{se}&B\rho_{ss}+C\rho_{ee}&P\rho_{sa}&T\rho_{sg}+U\rho_{es}\\
M^\ast\rho_{ae} & P^\ast\rho_{as}&D\rho_{aa}+E\rho_{ee}&Q\rho_{ag}+V\rho_{ea}\\
L^\ast\rho_{ge} & T^\ast\rho_{gs}+U^\ast\rho_{se}&Q^\ast\rho_{ga}+V^\ast\rho_{ae}& \rho_{gg}+F\rho_{ss}+\\
~ & ~& ~& G\rho_{aa}+H\rho_{ee}\\
\end{array} \right),\nonumber\\
\label{eq:channel}
\end{eqnarray}
where the functions $A, B,$ etc.~are given in Appendix \ref{ap:A}.

\section{Correlated noise \label{sec:D}}

In the regime, $k_{0}r_{12} \ll 1$, we expect the collective decoherence of the pair of qubits, whose strength is given by $\Gamma_{12}$, to produce correlated noise on the qubit pair. This would mean that the probabilities for simultaneous errors on the two qubits will not factorize: $p(x_1,x_2)\neq p(x_1)p(x_2)$. In general, a dynamical map $\mathcal{E}_{1,2,\cdots,n}$ acting on $n$ qubits as $\rho_{1,2,\cdots,n}^\prime = \mathcal{E}_{1,2,\cdots,n} \left(\rho_{1,2,\cdots,n}\right)$ produces correlated noise. The noise is  uncorrelated only if the action of the map is independent on each qubit; i.e,
\begin{equation}
	\mathcal{E}_{1,2,\cdots,n} = \mathcal{E}_1\otimes\mathcal{E}_2\otimes\cdots \otimes \mathcal{E}_n.
\label{eq:indep}
\end{equation}
If the  initial state  is in the  product form  $\rho_{1,2,\cdots,n} =
\rho_1  \otimes \rho_2 \otimes  \cdots \otimes  \rho_n$, noise  of the
type  in Eq.  (\ref{eq:indep})  will preserve  the  product form.  Any
departure  of  $\mathcal{E}_{1,2,\cdots,n}(\rho_{1,2,\cdots,n})$  from
the product  form leads to correlated  noise which in  turn will give rise to
nonclassical  correlations in quantum states  $\rho_{1,2,\cdots,n}$ characterized  by quantum discord. 
%This is shown for the case of two qubits initially in the product state in Appendix \ref{ap:dis}. \bla
%steering,  non-vanishing   non-locality  etc [REFS], between the qubits it acts on. 
For the two qubit case, the noise is uncorrelated if we can write the process matrix in the form
\begin{equation}
	\mathcal{E}(\rho) = \sum_{jk,lm} \chi_{jk}\chi_{lm}(F_j\otimes F_l)\rho  (F_k\otimes F_m),
	\label{eq:tensor}
\end{equation}
From Eq. (\ref{eq:tensor}) we see that for uncorrelated noise, the process matrix $\chi$ has the product form $\chi_{1}\otimes \chi_{2}$.

%\subsection{Quantifying the correlation in the noise}

We can quantify the degree of correlation in the noise on the two qubit system by defining a measure for the same as
\begin{equation}
	\label{eqn:cor}
	  {\mathcal D} = \| \chi - (\chi_{1} \otimes \chi_{2}) \|, 
\end{equation}
where $\|  \cdot \|$  denotes an appropriate  distance measure  in the
space of two qubit process matrices $\chi$. Since the process matrices
are normal, we  can use the trace distance as  an appropriate distance
measure~ \cite{NC00}.  The trace distance between  two normal matrices
$\rho_A$ and $\rho_B$ is defined as
\[ \|\rho_A-\rho_B\| \equiv \frac{1}{2} \sqrt{(\rho_A-\rho_B)^{\dagger}(\rho_A-\rho_B)} = \frac{1}{2} \sum_{j=1}^{d} |\lambda_{j}|, \]
where $\lambda_{j}$ are the eigenvalues (not necessarily real or positive) of the normal matrix $\rho_A-\rho_B$. 

  Note   that  any   other  suitable   measure  of   distance  or
  distinguishability that captures the above idea would also work, for
  example,  quantum mutual  information defined  via quantum  relative
  entropy:
\begin{eqnarray}
\mathcal{D}^\ast &\equiv& S(\rho_{AB}||\rho_A\otimes\rho_B) \nonumber\\
   &=& S(\rho_A)+S(\rho_B)-S(\rho_{AB}),
\label{eq:qre}
\end{eqnarray}
where  $S(\cdot)$  is von  Neumann  entropy  and $S(\cdot||\cdot)$  is
quantum  relative entropy.  The  quantity $\mathcal{D}^\ast$  vanishes
precisely   if  $\rho_{AB}$   has  the   product  form.   In  general,
$S(\rho_1||\rho_2)$  may diverge,  which would  happen if  there is  a
non-vanishing overlap between  the support of $\rho_1$  and the kernel
of  $\rho_2$.  For $\mathcal{D}^\ast$,  this  problem  does not  arise
because the two arguments in the definition have identical support.

\section{Characterization of $\mathcal{E}^{\rm 2AD}$ via 5-qubit QECCD}
\label{sec:characterization}
\begin{figure}
\includegraphics[width=0.480\textwidth]{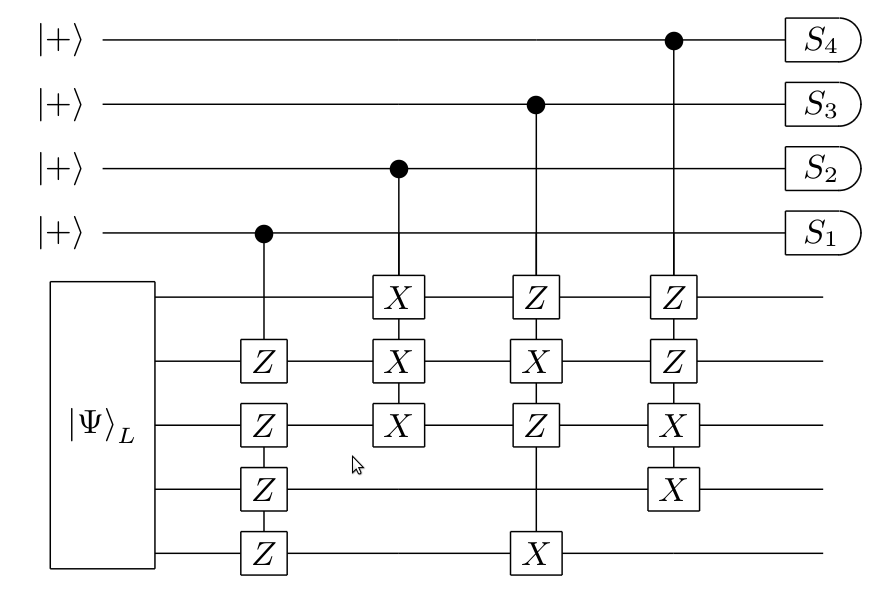}
\caption{Circuit to measure the stabilizer generators for our 5-qubit
  code using only one and two qubit interaction. The first four inputs
  to   the   circuit   are   the   ancilla   qubits   in   the   state
  $\ket{+}=\frac{\ket{0}+\ket{1}}{\sqrt{2}}$.   The  last  five  input
  qubits  correspond  to  the  QEC   code  initialized  in  the  state
  $\ket{\Psi}_L$.   The  boxes  represent  the  controlled  operations
  ($X,Y,Z$) with the control being the  filled circle at the other end
  of  the wire.   The  boxes $S_j$  represent measurement  in the
    computational basis.}
\label{fig:QC}
\end{figure}

We use  the the  $[[5,1]]$ QEC code  introduced in  Ref. \cite{OSB15},
with two logical basis states,
\begin{eqnarray} 
	|0_L\rangle&=&\frac{1}{2\sqrt{2}} (|00000\rangle + |00110\rangle + |01001\rangle - |01111\rangle \nonumber\\
	&& \qquad -|10011\rangle +|10101\rangle +|11010\rangle +|11100\rangle)\nonumber\\
	|1_L\rangle &=& XXXXX|0_L\rangle,
	\label{eq:5qecc}
\end{eqnarray}
to reconstruct ${\mathcal{E}}^{\rm 2AD}$ using QECCD. The code is capable of correcting arbitrary errors on the first two qubits which we consider as the primary system ${\mathbf P}$. The QEC code states are represented in the computational basis and the stabilizers generators of the code are the mutually commuting operators, $IZZZZ, ~ XXXII,~ ZXZIX,$ and $ ZZXXI$. An arbitrary logical state of the QEC code has the form
\begin{equation}
	\label{eq:logical1} |
	\Psi_L\rangle=\beta_{0}|0_L\rangle+ \beta_{1}|1_L\rangle, \qquad |\beta_{0}|^{2} +  |\beta_{1}|^{2} = 1.
\end{equation}
The register of five qubits is initialized in a logical state as given
in  Eq.~(\ref{eq:logical1})  and  the  primary  system  \textbf{P}  is
subjected to $\mathcal{E}^{\rm 2AD}$  noise, while the remaining three
ancilla qubits  are assumed to  be noise  free. The initial  and final
states of the five qubit register are therefore given by
\[\rho_c=|\Psi_L\rangle\langle\Psi_L|, \qquad {\rm and} \qquad \rho_c^{2\rm AD}=\left(\mathcal{E}^{\rm 2AD}\otimes I^{\otimes 3}\right)\rho_c , \] respectively.
The sixteen  operators corresponding  to strings  of length  two taken
from the set $\{I,X,Y,Z\}\otimes\{I,X,Y,Z\}$  form the error basis and
the  corresponding  syndromes  obtained through  measurements  of  the
stabilizers are given in Table~\ref{tab:synd}. Fig. \ref{fig:QC} shows
the quantum  circuit to implement  the stabilizer measurement  for the
5-qubit QEC code in Eq.  (\ref{eq:5qecc}) using only one and two qubit
interactions \cite{GPS+07}.
\begin{table}[!htb]
\begin{center}
\begin{tabular}{|l|c|c|c|c|c|c|c|c|c|c|c|c|c|c|c|c|}\hline
	Error &$I I $& $XI$&$ YI $&$ZI$&$ IX$&$ IY $ &$ IZ $&$XX$ \\ \hline
	$IZZZZ$ & + & + & + & + & -- & -- & + & -- \\ \hline
	$XXXII$ & + & + & -- & -- & + & -- & -- & + \\ \hline
	$ZXZIX$ & + & -- & -- & + & + & -- & -- & -- \\ \hline
	$ZZXXI$ & + & -- & -- & + & -- & -- & + & + \\ \hline
	\hline
	Error &$XY $& $XZ$&$ YX $&$YY$&$YZ$&$ ZX $ &$ ZY $&$ZZ$ \\ \hline
	$IZZZZ$ & -- & + & -- & -- & + & -- & -- & + \\ \hline
	$XXXII$ & -- & -- & -- & + & + & -- & + & + \\ \hline
	$ZXZIX$ & + & + & -- & + & + & + & -- & -- \\ \hline
	$ZZXXI$ & + & -- & + & + & -- & -- & + & + \\ \hline
\end{tabular}
\end{center}
\caption{The error syndromes (patterns of measurement outcomes of the stabilizer operators) corresponding to errors on $\textbf{P}$. "+" represents +1 measurement out come upon the measurement of stabilizer $S_j$ on $\rho_c^{2\rm AD}$, while the "-" stands for measurement
out come being -1.}
\label{tab:synd}
\end{table}

\subsection{Diagonal terms of $\chi^{2\textrm{AD}}$}
As mentioned in  Sec.~\ref{sec:diag} one can measure  all the diagonal
terms of the  $\chi$ matrix in a single  apparatus configuration since
all the stabilizers  commute with each other. Knowing  the error model
and using Eq.~(\ref{eq:channel}) that gives  the time evolution of the
system  ${\mathbf P}$,  we  can compute  the  expected statistics  for
obtaining each of the syndromes in Table.~\ref{tab:synd}. This in turn
gives  the diagonal  elements of  the  process matrix  of the  channel
$\mathcal{E}^{2AD}$ as in Eq.~(\ref{eq:diag}).

Equations.~(\ref{eq:inistate}) and (\ref{eq:channel}) give the states of ${\mathbf P}$ before and after the action of the noise.
The Ref. \cite{OSB15/1}  provides the operator sum-difference representation of  $\mathcal{E}^{2AD}$ with Kraus-like operators 
$\mathcal{K}_j^\pm$ according to which the action of the noise on the $\bf{P}$ of  $|\Psi_L\rangle$ is
\bea
\rho_c^{2\rm AD}&=&\sum_j(\mathcal{K}_j^+\otimes\openone_2^{\otimes3})\rho_c(\mathcal{K}_j^+\otimes\openone_2^{\otimes3})^\dag\nonumber\\
&&~~~-(\mathcal{K}_j^-\otimes\openone_2^{\otimes3})\rho_c(\mathcal{K}_j^-\otimes\openone_2^{\otimes3})^\dag,
\eea
where $\openone_2$ is identity on qubit. Before applying the $\mathcal{E}^{2AD}$, $\bf{P}$ is transformed to the dressed state basis using the unitary $U_{\rm D} \otimes \openone^{\otimes 3}$, where 
\[ U_{\rm D} = \left( \begin{array}{cccc} 1 & 0 & 0 & 0 \\ 0 & 1/\sqrt{2} & 1/\sqrt{2} & 0 \\ 0 & 1/\sqrt{2} & - 1/\sqrt{2} & 0 \\ 0 & 0 & 0 & 1 \end{array} \right), \]
and back to $\{0,1\}$ basis after application of $\mathcal{E}^{2AD}$ using $(U_{\rm D} \otimes \openone^{\otimes 3})^{-1}$.
Alternatively, by inspection, one can write down the map corresponding to the noise in the matrix super-operator from of Eq.~(\ref{eq:Aform}). The matrix ${\mathcal A}$ corresponding to the $\mathcal{E}^{2\rm AD}$ is a $16 \times 16$ sparse matrix with the following entries in the diagonal: 
\[ (A,J,M,L,J^{*},B,P,T,M^{*},P^{*},D,Q,L^{*},T^{*},Q^{*},1), \]
where $A, B \ldots$ are given in Appendix~\ref{ap:A}. The remaining non-zero elements of ${\mathcal A}$ are ${\mathcal A}_{6,1} = C$, ${\mathcal A}_{8,2} = U$, ${\mathcal A}_{11,1} = E$, ${\mathcal A}_{12,3} = V$, ${\mathcal A}_{14,5} = U^{*}$, ${\mathcal A}_{15,9} = V^{*}$, ${\mathcal A}_{16,1} = H$, ${\mathcal A}_{16,6} = F$ and ${\mathcal A}_{16,11} = G$.
After transforming the initial logical state from Eq. (\ref{eq:logical1}) into the dressed state basis using the unitary $U_{\rm D} \otimes \openone^{\otimes 3}$, 
%where \[ U_{\rm D} = \left( \begin{array}{cccc} 1 & 0 & 0 & 0 \\ 0 & 1/\sqrt{2} & 1/\sqrt{2} & 0 \\ 0 & 1/\sqrt{2} & - 1/\sqrt{2} & 0 \\ 0 & 0 & 0 & 1 \end{array} \right), \]
we can apply the map ${\mathcal A} \otimes \openone_{4}^{\otimes 3}$ on the vectorized form of $|\Psi_{L} \rangle \langle \Psi_{L}|$. Here $\openone_{4}$ in the identity super-operator on each of the vectorized ancilla qubits. Reshuffling the result back into a density matrix we get the state of the register after the action of the noise which can be expanded in the error basis as
\[ \rho_{c}^{2\rm AD} = \sum_{lm}^{d^2-1}\chi_{l,m}  |\Psi_L^l\rangle\langle \Psi_L^m|. \]
%with the subscript reminding us that this is a 5 qubit density matrix. If we trace out the three ancilla qubits out of the density matrix above, we get back $\rho^{{\mathcal{E}}_{\rm AD}}$ from Eq.~(\ref{eq:channel}). 
Using Eq.~(\ref{eq:diag}), the diagonal elements of $\chi^{2\rm AD}$ are obtained as
\begin{eqnarray*}
	\chi_{II,II}&=&\frac{1}{16} \big[1+A+B+D + 2(\textrm{Re}(J)+ \textrm{Re}(L) \\
	&&+ \textrm{Re}(M)+ \textrm{Re}(P)+ \textrm{Re}(Q)+ \textrm{Re}(T)) \big],
\end{eqnarray*}
\begin{eqnarray*}
	\chi_{XI,XI}&=&\chi_{IX,IX}=\chi_{YZ,YZ}=\chi_{ZY,ZY} \\
	& = & \frac{1}{32}\big[C+E+F+G + 2(\textrm{Re}(U)- \textrm{Re}(V)\big],  
\end{eqnarray*}
\begin{eqnarray*}
	\chi_{YI,YI}&=&\chi_{IY,IY}=\chi_{XZ,XZ}=\chi_{ZX,ZX} \\
	&=&\frac{1}{32}\big[C+E+F+G - 2(\textrm{Re}(U)- \textrm{Re}(V)\big],
\end{eqnarray*}
\begin{eqnarray*}
	\chi_{ZI,ZI}&=\chi_{IZ,IZ}&=\frac{1}{16}\big[1+A-2\textrm{Re}(L)\big],\\
	\chi_{XX,XX}&=\chi_{YY,YY} &=\frac{1}{16}\big[B+D+H-2\textrm{Re}(P)\big], \\
	\chi_{XY,XY}&=\chi_{YX,YX}&= \frac{1}{16}H,
\end{eqnarray*}
\begin{eqnarray*}
	\chi_{ZZ,ZZ}&=&\frac{1}{16}\big[1+A+B+D - 2(\textrm{Re}(J)- \textrm{Re}(L) \\
	&& +\textrm{Re}(M)- \textrm{Re}(P)+ \textrm{Re}(Q)- \textrm{Re}(T))\big].
\end{eqnarray*}

The noise is symmetric on the two qubits i.e., action of the noise under the change of the label of the qubits is invariant.
This is reflected in the fact that under the interchange of the qubit-label the equations above remain the same.

\subsection{Off-diagonal terms of $\chi^{\textrm{2AD}}$} 

To find the off-diagonal terms of $\chi^{\textrm{2AD}}$ using  QECCD as described in Sec.~\ref{sec:offdiag}, we have to pre-process the state $\rho_{c}^{2\rm AD}$ with unitaries $U(a,b)$ described in Eq. (\ref{eq:uab}) with suitable $a$ and $b$ chosen from the error basis. These will however yield only either the real or imaginary part of the off-diagonal terms of the process matrix. To obtain the other missing part, we will have to do an additional toggling operation that exchanges the real and imaginary parts of $\chi^{\textrm{2AD}}$. The necessary toggling operators are constructed  following the heuristic described in the Sec. \ref{sec:tog}. 

To determine the real/imaginary part of $\chi_{l,m}$ not possible by applying $U(a,b)$, place equal number of the $l^{\rm th}$ and the $m^{\rm th}$ error elements in the opposite partitions created by the coefficients $(1+i)$ and $(1-i)$ as
\be
T^+=\frac{1}{\sqrt{2}}\left[ (1+i)\sum_lF_l\rho_cF_l +(1-i)\sum_mF_m\rho_cF_m \right]. 
\ee
For example to toggle the elements in the table corresponding to $U(II,IZ)$ in Appendix \ref{sec:U}, place the elements $II, IX, XI,XX, IY,YX, IZ, ZX$ and $IZ,IY, XZ,XY, YZ,YY, ZZ,ZY$ in the different partitions i.e.,
\begin{eqnarray*}
T^+_1&=&\frac{1+i}{\sqrt{2}}\big(\rho_c+IX\rho_c IX +XI\rho_c XI +  XX \rho_c XX \\
&& \quad  +YI\rho_c YI +YX\rho_c YX   +ZI\rho_c ZI+ZX\rho_c ZX\big)\\
&&  + \frac{1-i}{\sqrt{2}}\big(IZ\rho_c IZ+ IY\rho_c IY + XZ\rho_cXZ \\
&& \quad  +XY\rho_c XY  + YZ\rho_c YZ+YY\rho_c YY\\
&& \qquad \qquad +ZZ\rho_c ZZ  +ZY\rho_c ZY\big).
\end{eqnarray*}
Similarly, other operators $T^+_2,\dots,T^+_7$ can be constructed.

The tables in Appendix~\ref{sec:U} list the necessary unitary and toggling operations as well as the syndrome measurements that are to be performed to obtain both real and imaginary parts of all the non-zero off-diagonal terms of the process matrix $\mathcal{\chi}^{\rm 2AD}$.

Further, in the independent regime, $r_{12}/\lambda_{0} \gg 1$ the quantities $A$, $B$, $C$,  $D$,  $E$, $J$, $L$, $M$, $P$, $Q$, $T$, $U$ and $V$ all go to zero when $t \rightarrow \infty$, where as $F$, $G$, and $H$ become equal to 1.  All the diagonal and off-diagonal terms of the process matrix $\chi^{\textrm{2AD}}$  reach the value 1/16 asymptotically and the matrix factors out as follows
\be
^{\rm ind}\chi^{2\rm AD}_{\rm asy}=\frac{1}{16}\left(
\begin{array}{cccccccccccccccc}
1&0&0&-1\\
0&1&i&0\\
0&-i&1&0\\
-1&0&0&1
\end{array}
\right)\otimes
\left(
\begin{array}{cccccccccccccccc}
1&0&0&-1\\
0&1&i&0\\
0&-i&1&0\\
-1&0&0&1
\end{array}
\right).
\ee

In the collective regime where $r_{12}/\lambda_{0} \rightarrow 0$, from Eq.~(\ref{gamma}) we see that $\Gamma_{12} = \Gamma_{21} = \Gamma$ since ${\mathcal{F}}(0) = 2/3$. Then we see that $A$, $B$,  $C$, $E$,  $J$, $L$, $M$, $P$, $T$, $U$, and $V$ go to zero as $t\rightarrow \infty$. At the same time we have   $D$, $F$, $G$, and $H$ going to 1 while $Q\rightarrow e^{-i(\omega_0-\Omega_{12})t}$.  This means that 
\[ \chi_{II,II} = \chi_{ZZ,ZZ} = \frac{2+ \cos (\omega_{0} - \Omega_{12})}{16},\] 
are oscillatory functions in the collective regime with a frequency $\omega_0 - \Omega_{12}$. Note that the function ${\mathcal G}$ in Eq.~(\ref{eq:omega12}) goes to infinity when $r_{12}/\lambda_{0} \rightarrow 0$. So $\chi_{II,II}$ and $\chi_{ZZ,ZZ}$ are rapidly oscillating functions of time when $r_{12}$ is small. When the two qubits are very close together, we can average over the rapid oscillations and then  $\chi_{II,II}$ and $\chi_{ZZ,ZZ}$ take on the limiting value of $1/8$ along with the diagonal terms $\chi_{YZ,YZ}$ and $\chi_{XY,XY}$. The terms $\chi_{Z_2,Z_2}$,  $\chi_{XY,XY}$ and $\chi_{YY,YY}$ become $1/8$ while  rest of the diagonal terms go to $1/32$ as $t\rightarrow\infty$ and the process matrix has non-factorizable asymptotic form: 
\begin{widetext}
\be
^{\rm col}\chi^{2\rm AD}_{\rm asy}=
\frac{1}{32}\left(
\begin{array}{cccccccccccccccc}
 4 & 0 & 0 & -2& 0 & -2& 0 & 0 & 0 & 0 & -2& 0 & -2& 0 & 0 & 0 \\
 0 & 1&  i& 0 & 1& 0 & 0 & -1&  i& 0 & 0 &  -i& 0 & -1&  -i& 0 \\
 0 &  -i& 1& 0 &  -i& 0 & 0 &  i& 1& 0 & 0 & -1& 0 &  i& -1& 0 \\
 -2& 0 & 0 & 2& 0 & 0 & 0 & 0 & 0 & 0 & 0 & 0 & 2& 0 & 0 & -2\\
 0 & 1&  i& 0 & 1& 0 & 0 & -1&  i& 0 & 0 &  -i& 0 & -1&  -i& 0 \\
 -2& 0 & 0 & 0 & 0 & 4 &  2i& 0 & 0 &  2i& 0 & 0 & 0 & 0 & 0 & 2\\
 0 & 0 & 0 & 0 & 0 &  -2i& 2& 0 & 0 & 2&  2i& 0 & 0 & 0 & 0 & 0 \\
 0 & -1&  -i& 0 & -1& 0 & 0 & 1&  -i& 0 & 0 &  i& 0 & 1&  i& 0 \\
 0 &  -i& 1& 0 &  -i& 0 & 0 &  i& 1& 0 & 0 & -1& 0 &  i& -1& 0 \\
 0 & 0 & 0 & 0 & 0 &  -2i& 2& 0 & 0 & 2&  2i& 0 & 0 & 0 & 0 & 0 \\
 -2& 0 & 0 & 0 & 0 & 0 &  -2i& 0 & 0 &  -2i& 4 & 0 & 0 & 0 & 0 & 2\\
 0 &  i& -1& 0 &  i& 0 & 0 &  -i& -1& 0 & 0 & 1& 0 &  -i& 1& 0 \\
 -2& 0 & 0 & 2& 0 & 0 & 0 & 0 & 0 & 0 & 0 & 0 & 2& 0 & 0 & -2\\
 0 & -1&  -i& 0 & -1& 0 & 0 & 1&  -i& 0 & 0 &  i& 0 & 1&  i& 0 \\
 0 &  i& -1& 0 &  i& 0 & 0 &  -i& -1& 0 & 0 & 1& 0 &  -i& 1& 0 \\
 0 & 0 & 0 & -2& 0 & 2& 0 & 0 & 0 & 0 & 2& 0 & -2& 0 & 0 & 4 \\
\end{array}
\right)
\ee
\end{widetext}

\subsection{Noise correlation in the 2AD channel}

\begin{figure}[!htb]
\subfigure[]{
\label{fig:sp-t}
\includegraphics[width=0.50\textwidth]{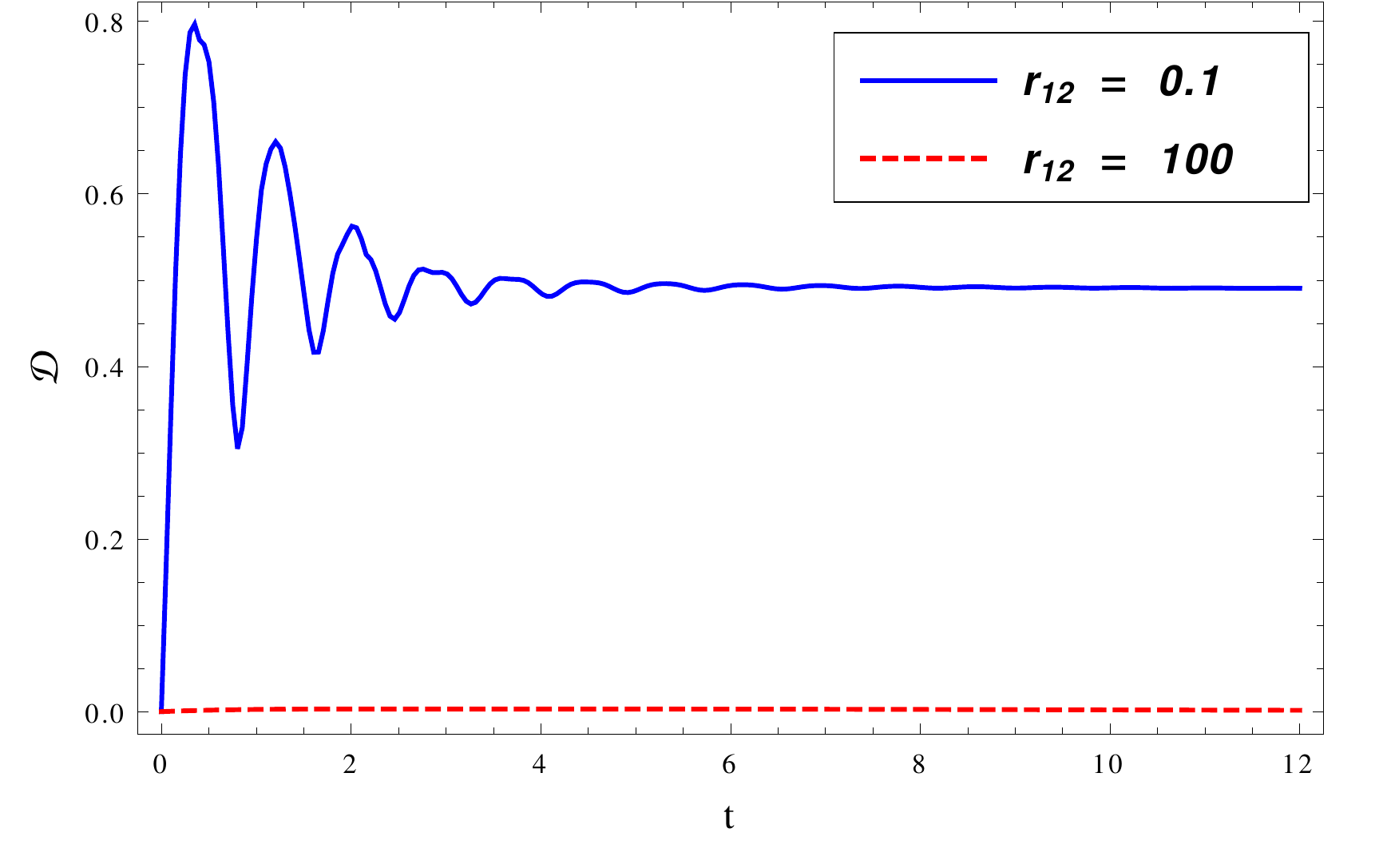}}
\subfigure[]{
\label{fig:sp-r}
\includegraphics[width=0.49\textwidth]{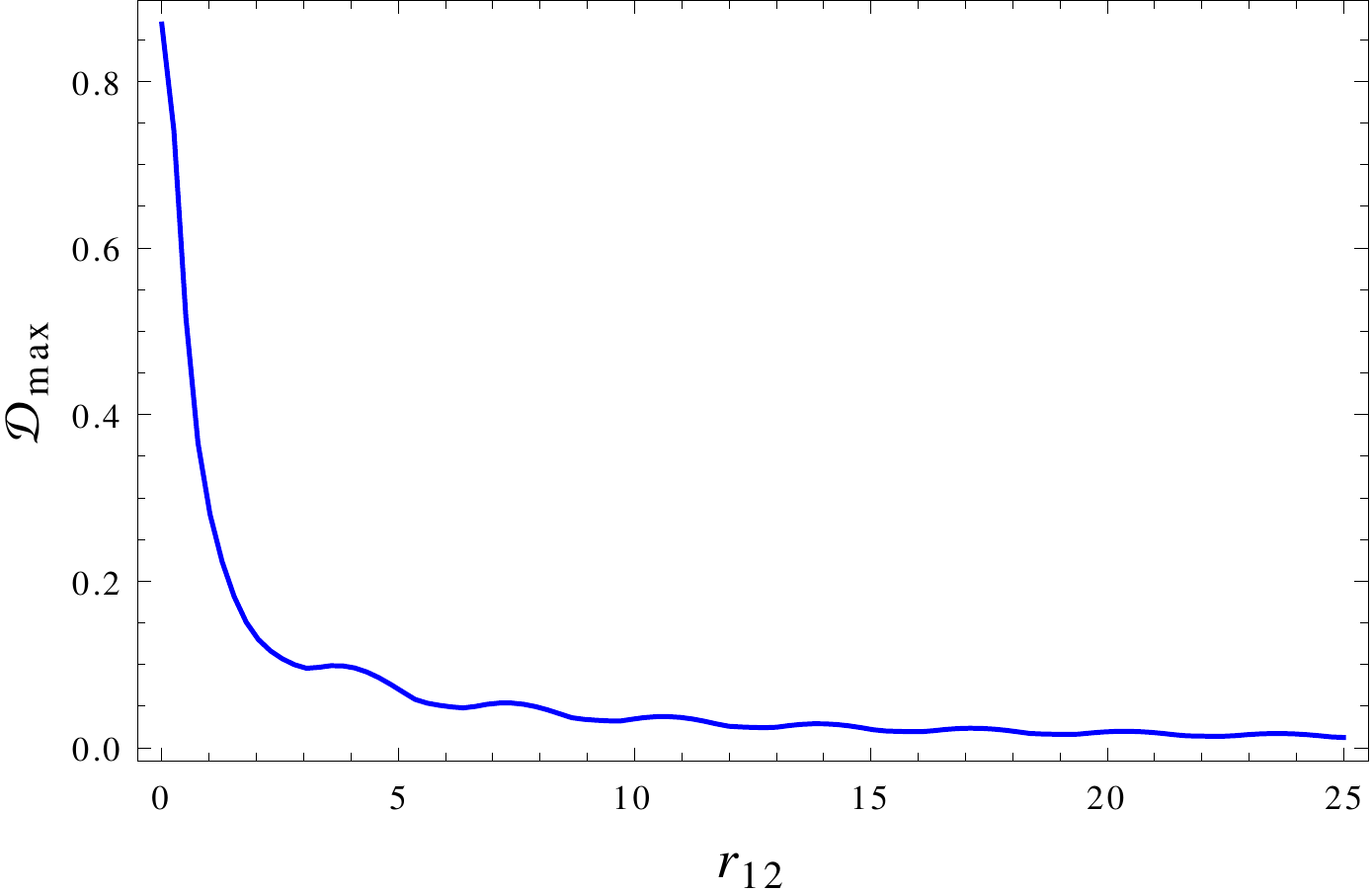}}
\caption{(a) $\mathcal{D}$  of Eq.(\ref{eqn:cor}) plotted  against $t$
  both in  the collective regime  (solid curve, $r_{12}=0.1$)  and the
  independent regime (dashed curve,  $r_{12}=100$). In the independent
  regime, there is no correlation between  the noise acting on the two
  qubits.   (b)   $\mathcal{D}_{\textrm{max}}$   is   the   value   of
  $\mathcal{D}$ maximized over  $t$ for each $r_{12}$ is  plotted as a
  function of $r_{12}$. We see that for large values of the inter-qubit
  separation, the correlation in the noise approaches zero. The values assigned to the 
  bath parameters are $\gamma_0=0.5$, $\omega_0=1$ and $k_0=1$ in the units where $\hbar=1$.}
\end{figure}

Since   the    coefficients   $A$,   $B$,   etc.    that   appear   in
Eq.~(\ref{eq:channel})   are  time   dependent,   the  dynamical   map
${\mathcal  E}^{\rm 2AD}$  and the  corresponding process  matrix also
have time  dependence. In Fig.~\ref{fig:sp-t}, the  time dependence of
the measure of correlation in the noise, ${\mathcal D}$ is plotted for
the cases where the two qubits are close by with $r_{12}=0.1$ in units
of  $\lambda_{0}$ and  where  the two  are far  apart  with $r_{12}  =
100$. We  see that  when the two  qubits are close  to each  other the
measure  of correlation  in the  noise is  finite indicating  that the
coherent  as  well as  incoherent  couplings  between the  two  qubits
mediated  by  the   bath  leads  to  correlated  errors   on  the  two
qubits.  When  $r_{12}$  is  large,  the  measure  of  correlation  is
identically zero at all times.  In Fig.~\ref{fig:sp-r}, the maximum of
${\mathcal D}$  as a  function of  time is  plotted against  the inter
qubit separation  $r_{12}$. We again  see that, as expected,  when the
separation   increases,    the   degree   of   correlation    in   the
$\mathcal{E}^{2\rm AD}$ noise decreases.

%Further, the induction of non-classical correlation in to the initial product state $\rho_{1,2}$ such as concurrence, non-locality, teleportation fidelity and discord by the ${\mathcal E}^{\rm 2AD}$ is been studied in Ref. \cite{CBS11}.

%\begin{figure}[!htb]
%\label{fig:sp-t}
%\includegraphics[width=0.50\textwidth]{d-t.pdf}
%\caption{$\mathcal{D}$ of Eq.(\ref{eqn:cor}) plotted against $t$ both in the collective regime (solid curve, $r_{12}=0.1$) and the %independent regime (dashed curve, $r_{12}=100$). In the independent regime, there is no correlation between the noise acting on the two qubits}
%\end{figure}

%\begin{figure}[!htb]
%\label{fig:sp-r}
%\includegraphics[width=0.50\textwidth]{dmax-r.pdf}
%\caption{$\mathcal{D}_{\textrm{max}}$  is the value of $\mathcal{D}$ maximized over $t$ for each $r_{12}$  is plotted as a function of       $r_{12}$. We see that for large values of the inter-qubit separation, the correlation in the noise approaches zero.}
%\end{figure}

\begin{figure}[!htb]
\includegraphics[width=0.49\textwidth]{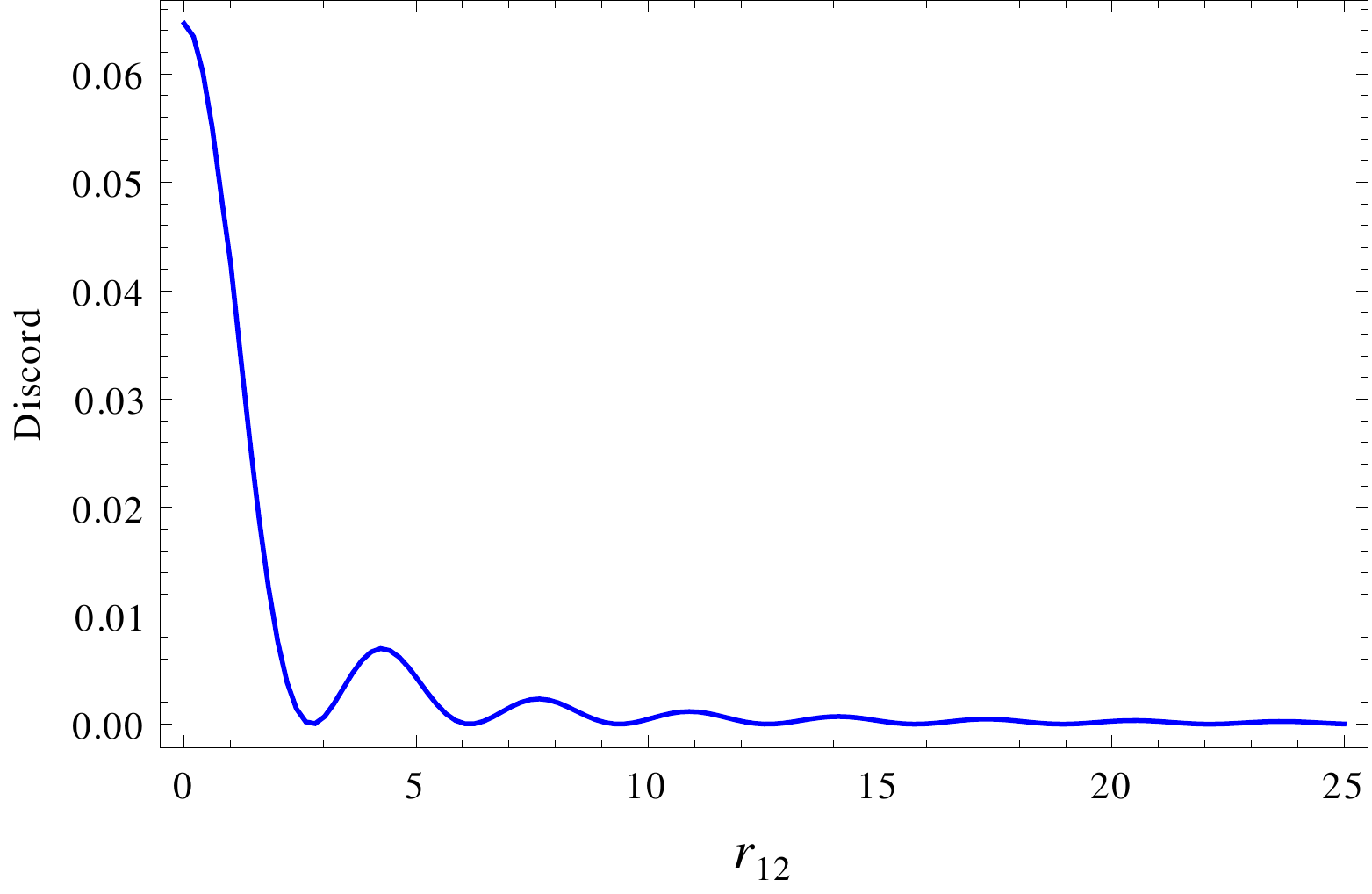}
\caption{The discord maximized  over $t$ for each  $r_{12}$ is plotted
  as a  function of  $r_{12}$.  We  see that for  large values  of the
  inter-qubit separation,  the discord approaches zero,  in accordance
  with  the asymptotically  vanishing $\mathcal{D}_{\rm max}$ in
  Fig. \ref{fig:sp-r}. The values assigned to the 
  bath parameters are $\gamma_0=0.5$, $\omega_0=1$ and $k_0=1$ in the units where $\hbar=1$.}
\label{fig:dis}
\end{figure}

In  view  of  Figure   \ref{fig:sp-r},  sufficient  far-spaced  qubits
subjected  to  the  2AD  noise   will  fail  to  produce  any  quantum
correlation  when   start  in  a   product  state.  However,   as  the
collective-noise regime is  approached, qubits start in  a product can
have non-classical correlation characterized  by quantum discord under
the $\mathcal{E}^{\rm 2AD}$ noise.

The correlated noise can generate non-classical correlations between the
qubits as characterized by  quantum discord %\cite{HV01,OZ02} 
even if the  qubits are in an initial product state.  In the  case of
$\mathcal{E}^{\rm2AD}$, as  the inter-qubit distance is  increased the
correlation  in the  noise  decreases resulting  in  the decrease  of
quantum  discord generated. Fig. \ref{fig:dis}  shows the  quantum discord  as
defined in Ref. \cite{DVB10} in  a two-qubit system (starting in the
product    state     $|0\rangle\otimes|0\rangle$)    evolving    under
$\mathcal{E}^{\rm  2AD}$   as  a  function  of   inter-qubit  distance
$r_{12}$.  The  value  of  quantum  discord  approaches  zero  as  the
correlation in  the $\mathcal{E}^{\rm  2AD}$ goes to  zero. 
%as  seen by comparing Fig. \ref{fig:dis} with Fig. \ref{fig:sp-r}\bla

%\begin{figure}[h]
%\subfigure[]{
%\label{fig:no-error}
%\includegraphics[width=0.45\textwidth]{stabi.pdf}}
%\caption{The stabilizer circuit for our 5-qubit code:
%square represents Z, round is X, controls in black filled round
%rounded, square is measurement in Hadamard basis and
%circle with + is preparation in Hadamard $|+\rangle$ state.}
%\end{figure}

\section{Discussion and conclusion \label{conclusion}}
In this work, we studied the dependence of the 2AD (2-qubit amplitude damping) channel on
inter-qubit separation, showing in particular that it becomes completely uncorrelated in the independent limit, but retains strong
correlation in the collective limit. Correlatedness of the noise is quantified in terms of the departure of the process matrix 
from the product form. Further, we presented a detailed QECCD protocol to completely characterize 2AD noise. The protocol can be 
easily extended for the characterization of both dissipative \cite{BRS10} and non-dissipative \cite{BRS10/1} dynamics due to 
squeezed-thermal bath, replacing the vacuum bath.

The protocol can be physically implemented on NMR systems\cite{SRM0}, linear optics with post selection \cite{KLM01}
and super conducting qubit system \cite{BKM+14} that are used for achieving fault tolerant quantum computation.

\bibliography{QECPT}

\appendix

\section{The coefficients appearing in $\rho_{t}$ \label{ap:A}}

\small{
\begin{eqnarray*}
	A&=& e^{-2\Gamma t}, \\
	B & = &  e^{-(\Gamma+\Gamma_{12})t},
\end{eqnarray*}
\begin{eqnarray*}
	C & = & \frac{\Gamma +\Gamma_{12}}{\Gamma -\Gamma_{12}} \big[ 1-e^{-(\Gamma -\Gamma_{12})t} \big] e^{-(\Gamma +\Gamma_{12})t}, \\
	D&=& e^{-(\Gamma-\Gamma_{12})t}, \\
	E & = & \frac{\Gamma -\Gamma_{12}}{\Gamma +\Gamma_{12}} \big[ 1-e^{-(\Gamma +\Gamma_{12})t} \big] e^{-(\Gamma -\Gamma_{12})t}, 
\end{eqnarray*}
\begin{eqnarray*}
	F& = & 1-e^{-(\Gamma +\Gamma_{12})t}, \\ 
	G&=&1-e^{-(\Gamma -\Gamma_{12})t}, 
\end{eqnarray*}}
\small{
\begin{eqnarray*}
	H & = & \frac{\Gamma +\Gamma_{12}}{2\Gamma} \bigg\{ 1-\frac{2}{\Gamma -\Gamma_{12}}  \bigg[ \frac{\Gamma +\Gamma_{12}}{2} \big( 1-e^{-(\Gamma -\Gamma_{12})t} \big)\\
	&& \qquad \qquad \qquad  + \, \frac{\Gamma -\Gamma_{12}}{2} \bigg] e^{-(\Gamma +\Gamma_{12})t}\bigg\} \\
	&+&\frac{\Gamma -\Gamma_{12}}{\Gamma +\Gamma_{12}} \bigg\{ 1-e^{-(\Gamma -\Gamma_{12})t} -\frac{\Gamma -\Gamma_{12}}{2\Gamma }(1-e^{-2\Gamma t})\bigg\}. 
\end{eqnarray*}}
\small{
\begin{eqnarray*}
	J & = & e^{-i(\omega_0 -\Omega_{12})t} e^{-(3\Gamma +\Gamma_{12})t/2}. \\
	L&=&e^{-i2\omega_0 t}e^{-\Gamma t}. \\
	M&=&e^{-i(\omega_0 +\Omega_{12})t} e^{-(3\Gamma -\Gamma_{12})t/2}. \\
	P & = & e^{-i2\Omega_{12}t}e^{-\Gamma t}. \\
	Q& = & e^{-i(\omega_0 -\Omega_{12})t} e^{-(\Gamma -\Gamma_{12})t/2}. \\
	T&=&e^{-i(\omega_0 +\Omega_{12})t} e^{-(\Gamma +\Gamma_{12})t/2}. \\
\end{eqnarray*}}
\small{
\begin{eqnarray*}
	U&=&\frac{\Gamma +\Gamma_{12}}{\Gamma ^2+4\Omega_{12}^2} e^{-i(\omega_0 + \Omega_{12})t} e^{-(\Gamma +\Gamma_{12})t/2} \\
	&& \times \Big\{ 2\Omega_{12} e^{-\Gamma t} \sin(2\Omega_{12} t) + \Gamma \big[ 1-e^{-\Gamma t}\cos(2\Omega_{12} t) \big] \Big\} \nonumber\\
	&&+~i~\frac{\Gamma +\Gamma_{12}}{\Gamma ^2+4\Omega_{12}^2} e^{-i(\omega_0 +\Omega_{12})t} e^{-(\Gamma +\Gamma_{12})t/2} \\
	&& \times \Big\{ 2\Omega_{12} \big[ 1-e^{-\Gamma t}\cos(2\Omega_{12} t) \big] -\Gamma e^{-\Gamma t} \sin(2\Omega_{12} t)  \Big\}.  
\end{eqnarray*}
\begin{eqnarray*}
V&=&~i~ \frac{\Gamma -\Gamma_{12}}{\Gamma ^2+4\Omega_{12}^2} e^{-i(\omega_0 -\Omega_{12})t} e^{-(\Gamma -\Gamma_{12})t/2} \\
&& \times \Big\{ 2\Omega_{12} \big[ 1-e^{-\Gamma t}\cos(2\Omega_{12} t) \big]-\Gamma e^{-\Gamma t} \sin(2\Omega_{12} t) \Big\} \\
&&-\frac{\Gamma -\Gamma_{12}}{\Gamma ^2+4\Omega_{12}^2} e^{-i(\omega_0 -\Omega_{12})t} e^{-(\Gamma -\Gamma_{12})t/2} \\
&& \times \Big\{ 2\Omega_{12} e^{-\Gamma t} \sin(2\Omega_{12} t)+\Gamma  \big[ 1-e^{-\Gamma t}\cos(2\Omega_{12} t) \big] \Big\}. 
\end{eqnarray*}
}

\section{Protocol to characterize the 2AD channel\label{sec:U}}

The  tables below  give the  unitary operators,  measurements and  the
toggling operators to  be used to pre-process the noisy  state and the
syndrome statistics that  are to be collected for  estimating the real
and  imaginary  parts of  the  off-diagonal  elements of  the  process
matrix. In  the tables  below the first  column indicates  the unitary
operator to  be used, the second  column indicates the syndrome  to be
measured and the third column lists the toggling operators (if any) to
be used. The  last column shows the element of  $\chi^{\rm 2 AD}$ that
is measured along with its value.

\begin{widetext}

\begin{center}

\begin{tabular}{ |l|c|l|l| } 
\hline
%$U(a,b)$ & $S$ & Toggling & $\textrm{Re}(\chi_{m,n})/\textrm{Im}(\chi_{m,n}) \phantom{\Big|}$ \\ \hline
\multirow{3}{4em}{$U(II,IZ)$} \hspace{5 mm}& $ II$ & None &$\textrm{Re}(\chi_{II,IZ})=$
 $\frac{1}{16}[ A-1+\textrm{Re}(J)+\textrm{Re}(M)-\textrm{Re}(Q)-\textrm{Re}(T)] \phantom{\Big|}$\\ \cline{3-4}
 & &$T^+_1$&$\textrm{Im}(\chi_{II,IZ})=$
 $\frac{1}{16}[2~\textrm{Im}(L)+\textrm{Im}(J)+\textrm{Im}(M)+\textrm{Im}(Q)+\textrm{Im}(T)] \phantom{\Big|}$\\ \cline{2-4}
 
 &$IX$ & $T^+_1$ &$\textrm{Re}(\chi_{IX,IY})=0 \phantom{\Big|}$ \\ \cline{3-4}
 & & None &$\textrm{Im}(\chi_{IX,IY})=\frac{1}{32}[C+E+F+G+2\textrm{Re}(U)-2\textrm{Re}(V)] \phantom{\Big|}$ \\ \cline{2-4}

 &$XI$ & None &$\textrm{Re}(\chi_{XI,XZ})=\frac{1}{32}(C+E-F-G) \phantom{\Big|}$ \\ \cline{3-4}
 & & $T^+_1$ &$\textrm{Im}(\chi_{XI,XZ})=\frac{1}{16}[\textrm{Im}(U)-\textrm{Im}(V)] \phantom{\Big|}$ \\ \cline{2-4}

 &$XX$ & $T^+_1$ & $\textrm{Re}(\chi_{XX,XY})=0 \phantom{\Big|}$ \\ \cline{3-4}
 & & None & $\textrm{Im}(\chi_{XX,XY})=\frac{1}{16}H \phantom{\Big|}$ \\ \cline{2-4}

 &$IY $ & None & $\textrm{Re}(\chi_{IY,YZ})=\frac{1}{32}(C+E-F-G) \phantom{\Big|}$ \\ \cline{3-4}
 & & $T^+_1$ & $\textrm{Im}(\chi_{IY,YZ})=\frac{1}{16}[ \textrm{Im}(U)-\textrm{Im}(V)] \phantom{\Big|}$ \\ \cline{2-4}

 &$YX$ & $T^+_1$ & $\textrm{Re}(\chi_{YX,YY})=0 \phantom{\Big|}$ \\ \cline{3-4}
 & & None & $\textrm{Im}(\chi_{YX,YY})=\frac{1}{16}H \phantom{\Big|}$ \\ \cline{2-4}

 & $IZ $ & None &$\textrm{Re}\chi_{IZ,ZZ}=$
 $\frac{1}{16}[A-1-\textrm{Re}(J)-\textrm{Re}(M)+\textrm{Re}(Q)+\textrm{Re}(T)] \phantom{\Big|}$\\ \cline{3-4}
 & &$T^+_1$&$\textrm{Im}\chi_{IZ,ZZ}=$
 $\frac{1}{16} [\textrm{Im}(J)-2\textrm{Im}(L)+\textrm{Im}(M)+\textrm{Im}(Q)+\textrm{Im}(T)] \phantom{\Big|}$\\ \cline{2-4}

 &$ZX$ & $T^+_1$ & $\textrm{Re}(\chi_{ZX,ZY})=0 \phantom{\Big|}$ \\ \cline{3-4}
 & & None & $\textrm{Im}(\chi_{YX,YY})=\frac{1}{32}[C+E+F+G-2(\textrm{Re}(X)-\textrm{Re}(Y))] \phantom{\Big|}$ \\ \hline
 
 \end{tabular}
 
 \begin{tabular}{ |l|c|l|l| } 
\hline
 
\multirow{3}{4em}{$U(II,ZI)$} \hspace{5 mm}& $II$ & None &$\textrm{Re}(\chi_{II,ZI})=$
 $\frac{1}{16}[A-1+\textrm{Re}(J)+\textrm{Re}(M)-\textrm{Re}(Q)-\textrm{Re}(T)]\phantom{\Big|}$\\ \cline{3-4}
 & &$T^+_2$&$\textrm{Im}(\chi_{II,ZI})=$
 $\frac{1}{16}[2~\textrm{Im}(L)+\textrm{Im}(J)+\textrm{Im}(M)+\textrm{Im}(Q)+\textrm{Im}(T)] \phantom{\Big|}$\\ \cline{2-4}

 &$IX$ & None & $\textrm{Re}(\chi_{IX,ZX})=\frac{1}{32}(C+E-F-G) \phantom{\Big|}$ \\ \cline{3-4}
 & & $T^+_2$ & $\textrm{Im}[\chi_{IX,ZX})=\frac{1}{16}(\textrm{Im}(U)-\textrm{Im}(V)] \phantom{\Big|}$ \\ \cline{2-4}

 &$XI$ & $T^+_2$ &$\textrm{Re}(\chi_{XI,YI})=0 \phantom{\Big|}$ \\ \cline{3-4}
 & & None &$\textrm{Im}(\chi_{XI,YI})=\frac{1}{32}[C+E+F+G+2\textrm{Re}(U)+2\textrm{Re}(V)] \phantom{\Big|}$ \\ \cline{2-4}

 &$XX$ & $T^+_2$ & $\textrm{Re}(\chi_{XX,YX})=0\phantom{\Big|}$ \\ \cline{3-4}
 & & None & $\textrm{Im}(\chi_{XX,YX})=\frac{1}{16}H\phantom{\Big|}$ \\ \cline{2-4}

 &$XY$ & $T^+_2$ & $\textrm{Re}(\chi_{XY,YY})=0\phantom{\Big|}$ \\ \cline{3-4}
 & & None & $\textrm{Im}(\chi_{XY,YY})=\frac{1}{16}H\phantom{\Big|}$ \\ \hline

\end{tabular}
 
 \begin{tabular}{ |l|c|l|l| } 
\hline

\multirow{3}{4em}{$U(II,XX)$}\hspace{2 mm} &$II$ & None &$\textrm{Re}(\chi_{II,XX})=$
 $\frac{1}{16}[B-D+\textrm{Re}(J)-\textrm{Re}(M)-\textrm{Re}(Q)+\textrm{Re}(T)] \phantom{\Big|}$\\ \cline{3-4}
 & &$T^+_3$&$\textrm{Im}(\chi_{II,XX}) =$
 $\frac{1}{16}[2~\textrm{Im}(P)-\textrm{Im}(J)+\textrm{Im}(M)-\textrm{Im}(Q)+\textrm{Im}(T)] \phantom{\Big|}$\\ \cline{2-4}

 &$IX$ & None &$\textrm{Re}(\chi_{IX,XI})=\frac{1}{32}[C-E+F-G+2\textrm{Re}(U)+2\textrm{Re}(V)] \phantom{\Big|}$ \\ \cline{3-4}
 & & $T^+_3$ &$\textrm{Im}(\chi_{IX,XI})=0 \phantom{\Big|}$ \\ \cline{2-4}

 &$IY$ &$T^+_3$ &$\textrm{Re}(\chi_{IY,XZ})=\frac{1}{16}[ \textrm{Re}(U)+\textrm{Re}(V)] \phantom{\Big|}$ \\ \cline{3-4}
 & & None &$\textrm{Im}(\chi_{IY,XZ})=-\frac{1}{32}(C-E-F+G)\phantom{\Big|}$ \\ \cline{2-4}

 &$YI$ & $T^+_3$ &$\textrm{Re}(\chi_{YI,ZX})=\frac{1}{16}[\textrm{Re}(X)+\textrm{Re}(Y)] \phantom{\Big|}$ \\ \cline{3-4}
 & & None &$\textrm{Im}(\chi_{YI,ZX})=-\frac{1}{32}(C-E-F+G) \phantom{\Big|}$ \\ \cline{2-4}

 & $YY$ & None &$\textrm{Re}(\chi_{YY,ZZ})=$
 $-\frac{1}{16}[-B+D+\textrm{Re}(J)-\textrm{Re}(M)-\textrm{Re}(Q)+\textrm{Re}(T)] \phantom{\Big|}$\\ \cline{3-4}
 & & $T^+_3$ &$\textrm{Im}(\chi_{YY,ZZ})=$
 $-\frac{1}{16}[2~\textrm{Im}(P)+\textrm{Im}(J)-\textrm{Im}(M)+\textrm{Im}(Q)-\textrm{Im}(T)] \phantom{\Big|}$\\ \cline{2-4}

 &$YZ$ & None &$\textrm{Re}(\chi_{YZ,ZY})=\frac{1}{32}[C-E+F-G-2\textrm{Re}(U)+2\textrm{Re}(V) ]\phantom{\Big|}$ \\ \cline{3-4}
 & &$T^+_3$ &$\textrm{Im}(\chi_{YZ,ZY})=0 \phantom{\Big|}$ \\ \hline

 \end{tabular}
 
\begin{tabular}{ |l|c|l|l| } 
\hline

\multirow{3}{4em}{$U(II,YY)$} \hspace{5 mm}&$II$ & None &$\textrm{Re}(\chi_{II,YY})=$
 $\frac{1}{16}[B-D+\textrm{Re}(J)-\textrm{Re}(M)-\textrm{Re}(Q)+\textrm{Re}(T)] \phantom{\Big|}$\\ \cline{3-4}
 & &$T^+_4$&$\textrm{Im}(\chi_{II,YY})=$
 $\frac{1}{16}[2~\textrm{Im}(P)-\textrm{Im}(J)+\textrm{Im}(M)-\textrm{Im}(Q)+\textrm{Im}(T)]\phantom{\Big|}$\\ \cline{2-4}

 &$IX$ & $T^+_4$ &$\textrm{Re}(\chi_{IX,YZ})=\frac{1}{16}[\textrm{Re}(U)+\textrm{Re}(V)] \phantom{\Big|}$ \\ \cline{3-4}
 & & None &$\textrm{Im}(\chi_{IX,YZ})=-\frac{1}{32}(C-E-F+G)\phantom{\Big|}$ \\ \cline{2-4}

 &$IY$ & None &$\textrm{Re}(\chi_{IY,YI})=\frac{1}{32}[ C-E+F-G+2\textrm{Re}(U)+2\textrm{Re}(V)] \phantom{\Big|}$ \\ \cline{3-4}
 & & $T^+_4$ &$\textrm{Im}(\chi_{IY,YI})=0\phantom{\Big|}$ \\ \cline{2-4}

 &$XI$ &$T^+_4$ &$\textrm{Re}(\chi_{XI,ZY})=\frac{1}{16}[\textrm{Im}(X)+\textrm{Im}(Y)] \phantom{\Big|}$ \\ \cline{3-4}
 & & None &$\textrm{Im}(\chi_{XI,ZY})=-\frac{1}{32}(C-E-F+G)\phantom{\Big|}$ \\ \cline{2-4}

 & $XX$ & None &$\textrm{Re}(\chi_{XX,ZZ})=$
 $-\frac{1}{16}[-B+D+\textrm{Re}(J)-\textrm{Re}(M)-\textrm{Re}(Q)+\textrm{Re}(T)]\phantom{\Big|}$\\ \cline{3-4}
 & & $T^+_4$ &$\textrm{Im}(\chi_{XX,ZZ})=$
 $-\frac{1}{16}[2~\textrm{Im}(P)+J\textrm{Im}(J)-\textrm{Im}(M)+\textrm{Im}(Q)-\textrm{Im}(T)] \phantom{\Big|}$\\ \cline{2-4}

 &$XZ$ & None &$\textrm{Re}(\chi_{XZ,ZX})=\frac{1}{32}[C-E+F-G-2\textrm{Re}(U)+2\textrm{Re}(V)] \phantom{\Big|}$ \\ \cline{3-4}
 & & $T^+_4$ &$\textrm{Im}(\chi_{XZ,ZX})=0\phantom{\Big|}$ \\ \hline
\end{tabular}

\begin{tabular}{ |l|c|l|l| } 
\hline
%$U(a,b)$ & $S$ & $Toggling$ & $\textrm{Re}(\chi_{m,n})/\textrm{Im}(\chi_{m,n})$ \\ \hline
\multirow{3}{4em}{$U(II,ZZ)$} \hspace{5 mm}& $II$ & None &$\textrm{Re}(\chi_{II,ZZ})=$
 $\frac{1}{16}[1+A-B-D+2\textrm{Re}(L)-2\textrm{Re}(P)] \phantom{\Big|}$\\ \cline{3-4}
 & &$T^+_5$&$\textrm{Im}(\chi_{II,ZZ})=$
 $\frac{1}{16}[2~\textrm{Im}(J)+2\textrm{Im}(M)-2\textrm{Im}(Q)-2\textrm{Im}(T)] \phantom{\Big|}$\\ \cline{2-4}
 
 &$IX$ & $T^+_5$ &$\textrm{Re}(\chi_{IX,ZY})=-\frac{1}{16}[\textrm{Im}(U)-\textrm{Im}(V)] \phantom{\Big|}$ \\ \cline{3-4}
 & & None &$\textrm{Im}(\chi_{IX,ZY})=\frac{1}{32}(C+E-F-G) \phantom{\Big|}$ \\ \cline{2-4}
 
 &$IY$ & $T^+_5$ &$\textrm{Re}(\chi_{IY,ZX})=\frac{1}{16}[\textrm{Im}(U)-\textrm{Im}(V)] \phantom{\Big|}$ \\ \cline{3-4}
 & & None &$\textrm{Im}(\chi_{IY,ZX})=-\frac{1}{32}(C+E-F-G) \phantom{\Big|}$ \\ \cline{2-4}
 
 &$IZ$ & None &$\textrm{Re}(\chi_{IZ,ZI})=\frac{1}{16}[1+A-2\textrm{Re}(L)]\phantom{\Big|}$ \\ \cline{3-4}
 & & $T^+_5$ &$\textrm{Im}(\chi_{IZ,ZI})=0 \phantom{\Big|}$ \\ \cline{2-4}

 &$XI$ & $T^+_5$ &$\textrm{Re}(\chi_{XI,YZ})=-\frac{1}{16}[\textrm{Im}(U)-\textrm{Im}(V)]\phantom{\Big|}$ \\ \cline{3-4}
 & & None &$\textrm{Im}(\chi_{XI,YZ})=\frac{1}{32}(C+E-F-G)\phantom{\Big|}$ \\ \cline{2-4}

 &$YI$ & $T^+_5$ &$\textrm{Re}(\chi_{YI,XZ})=-\frac{1}{16}[\textrm{Im}(U)+\textrm{Im}(V)]\phantom{\Big|}$ \\ \cline{3-4}
 & & None &$\textrm{Im}(\chi_{YI,XZ})=-\frac{1}{32}(C+E-F-G)\phantom{\Big|}$ \\ \cline{2-4}
 
 &$XX$ & None &$\textrm{Re}(\chi_{XX,YY})=\frac{1}{16}[B+D-H-2\textrm{Re}(P)] \phantom{\Big|}$ \\ \cline{3-4}
 & & $T^+_5$ &$\textrm{Im}(\chi_{XX,YY})=0\phantom{\Big|}$ \\ \cline{2-4}
 
 &$XY$ & None &$\textrm{Re}(\chi_{XY,YX})=\frac{1}{16}H\phantom{\Big|}$ \\ \cline{3-4}
 & & $T^+_5$ &$\textrm{Im}(\chi_{XY,YX})=0\phantom{\Big|}$ \\ \cline{1-4}
 
 \end{tabular}
 
\begin{tabular}{ |l|c|l|l| } 
\hline

\multirow{3}{4em}{$U(IX,YI)$} \hspace{5 mm}&$ II$ & None &$\textrm{Re}(\chi_{IX,YI})=0 \phantom{\Big|}$ \\ \cline{3-4}
 & & $T^+_6$&$\textrm{Im}(\chi_{IX,YI})=\frac{1}{32}[C-E+F-G+2\textrm{Im}(U)+2\textrm{Im}(V)] \phantom{\Big|}$ \\ \cline{2-4}
 
 &$XI$ & $T^+_6$ &$\textrm{Re}(\chi_{XX,ZI})=$
 $-\frac{1}{16}[\textrm{Re}(J)-\textrm{Re}(M)+\textrm{Re}(Q)-\textrm{Re}(T)]\phantom{\Big|}$\\ \cline{3-4}
 & & None &$\textrm{Im}(\chi_{XX,ZI}=$
 $-\frac{1}{16}[\textrm{Im}(J)-\textrm{Im}(M)+\textrm{Im}(Q)-\textrm{Im}(T)] \phantom{\Big|}$\\ \cline{2-4}

 &$XX$ & $T^+_6$ &$\textrm{Re}(\chi_{XI,ZX})=-\frac{1}{32}(C-E-F+G)\phantom{\Big|}$ \\ \cline{3-4}
 & & None &$\textrm{Im}(\chi_{XI,ZX})=-\frac{1}{16}[\textrm{Im}(U)+\textrm{Im}(V)] \phantom{\Big|}$ \\ \cline{2-4}
 
 &$XY$ & $T^+_6$ &$\textrm{Re}(\chi_{XZ,ZY})=0\phantom{\Big|}$ \\ \cline{3-4}
 & & None &$\textrm{Im}(\chi_{XZ,ZY})=-\frac{1}{32}[C+E+F+G-2\textrm{Im}(U)+2\textrm{Im}(V)]\phantom{\Big|}$ \\ \cline{2-4}
 
 &$YY$ & $T^+_6$ &$\textrm{Re}(\chi_{IY,YZ})=-\frac{1}{32}(C-E-F+G) \phantom{\Big|}$ \\ \cline{3-4}
 & & None &$\textrm{Im}(\chi_{IY,YZ})=-\frac{1}{16}[\textrm{Im}(U)+\textrm{Im}(V)] \phantom{\Big|}$ \\ \cline{2-4}
 
 &$YZ$ & $T^+_6$ &$\textrm{Re}(\chi_{YY,IZ})=$
 $-\frac{1}{16}[\textrm{Re}(J)-\textrm{Re}(M)+\textrm{Re}(Q)-\textrm{Re}(T)] \phantom{\Big|}$\\ \cline{3-4}
 & & None &$\textrm{Im}(\chi_{YY,IZ}=$
 $-\frac{1}{16}[-\textrm{Im}(J)+\textrm{Im}(M)+\textrm{Im}(Q)-\textrm{Im}(T)]\phantom{\Big|}$\\ \cline{1-4}

\end{tabular}
  \vspace{2 mm}

\begin{tabular}{ |l|c|l|l| } 
\hline

\multirow{3}{4em}{$U(IY,XI)$} \hspace{5 mm}&$II$ & None &$\textrm{Re}(\chi_{IY,XI})=0 \phantom{\Big|}$ \\ \cline{3-4}
 & &$T^+_7$ &$\textrm{Im}(\chi_{IY,XI})=-\frac{1}{32}[C-E+F-G+2\textrm{Re}(U)+2\textrm{Re}(V)] \phantom{\Big|}$ \\ \cline{2-4}
 
 &$IX$ &$T^+_7$ &$\textrm{Re}(\chi_{IZ,XX})=$
 $-\frac{1}{16}[\textrm{Re}(J)-\textrm{Re}(M)+\textrm{Re}(Q)-\textrm{Re}(T)] \phantom{\Big|}$\\ \cline{3-4}
 & & None &$\textrm{Im}(\chi_{IZ,XX}=$
 $-\frac{1}{16}[-\textrm{Im}(J)+\textrm{Im}(M)+\textrm{Im}(Q)-\textrm{Im}(T)] \phantom{\Big|}$\\ \cline{2-4}
 
 &$IZ$ & $T^+_7$ &$\textrm{Re}(\chi_{IX,XZ})=\frac{1}{32}(C-E-F+G) \phantom{\Big|}$ \\ \cline{3-4}
 & & None &$\textrm{Im}(\chi_{IX,XZ})=\frac{1}{16}[\textrm{Im}(U)+\textrm{Im}(V)] \phantom{\Big|}$ \\ \cline{2-4}
 
 &$YI$ &$T^+_7$ &$\textrm{Re}(\chi_{YY,ZI})=$
 $\frac{1}{16}[\textrm{Re}(J)-\textrm{Re}(M)+\textrm{Re}(Q)-\textrm{Re}(T)] \phantom{\Big|}$\\ \cline{3-4}
 & & None &$\textrm{Im}(\chi_{YY,ZI})=$
 $\frac{1}{16}[\textrm{Im}(J)-\textrm{Im}(M)+\textrm{Im}(Q)-\textrm{Im}(T)]\phantom{\Big|}$\\ \cline{2-4}

 &$YX$ & None &$\textrm{Re}(\chi_{YZ,ZX})=0 \phantom{\Big|}$ \\ \cline{3-4}
 & & $T^+_7$ &$\textrm{Im}(\chi_{YZ,ZX})=\frac{1}{32}[C-E+F-G-2\textrm{Re}(U)-2\textrm{Re}(V)]\phantom{\Big|}$ \\ \cline{2-4}
 
 &$YY$ & $T^+_7$ &$\textrm{Re}(\chi_{YI,ZY})=\frac{1}{32}(C-E-F+G) \phantom{\Big|}$ \\ \cline{3-4}
 & & None &$\textrm{Im}(\chi_{YI,ZY})=\frac{1}{16}[\textrm{Im}(U)+\textrm{Im}(V)] \phantom{\Big|}$ \\ \cline{1-4}
 
 \end{tabular}
 \vspace{2 mm}
 
\begin{tabular}{ |l|c|l|l| } 
\hline

\multirow{3}{4em}{$U(IY,ZY)$} \hspace{5 mm}&$II$ & None &$\textrm{Re}(\chi_{IY,ZY})=\frac{1}{32}(C+E-F-G) \phantom{\Big|}$ \\ \cline{3-4}
 & & $T^+_8$ &$\textrm{Im}(\chi_{IY,ZY})=\frac{1}{16}[\textrm{Im}(U)-\textrm{Im}(V)] \phantom{\Big|}$ \\ \cline{2-4}

 & $IX$ & None &$\textrm{Re}(\chi_{IZ,ZZ})=$
 $-\frac{1}{16}[A-1-\textrm{Re}(J)-\textrm{Re}(M)+\textrm{Re}(Q)+\textrm{Re}(T)] \phantom{\Big|}$\\ \cline{3-4}
 & &$T^+_8$&$\textrm{Im}(\chi_{IZ,ZZ})=$
 $-\frac{1}{16}[-2~\textrm{Im}(L)+\textrm{Im}(J)+\textrm{Im}(M)+\textrm{Im}(Q)+\textrm{Im}(T)] \phantom{\Big|}$\\ \cline{2-4}

 &$XX$ & $T^+_8$ &$\textrm{Re}(\chi_{XZ,YZ})=0\phantom{\Big|}$ \\ \cline{3-4}
 & & None &$\textrm{Im}(\chi_{XZ,YZ})=-\frac{1}{32}[C+E+F+G-2\textrm{Re}(U)+2\textrm{Re}(V)] \phantom{\Big|}$ \\ \cline{2-4}

\hline
\end{tabular}

\end{center}
\end{widetext}

\end{document}